\documentclass[aps,prd,twocolumn,groupedaddress,amssymb,eqsecnum,showpacs,
nofootinbib]{revtex4}
\usepackage{graphicx}
\usepackage{bm}

\begin{document}

\preprint{UMHEP-463,CMU-HEP-06-09}

\title{The energy-momentum tensor for an effective initial state and the renormalization of gravity}

\author{Hael Collins}
\email{hael@physics.umass.edu}
\affiliation{Department of Physics, University of Massachusetts, 
Amherst MA\ \ 01003}
\author{R.~Holman}
\email{rh4a@andrew.cmu.edu}
\affiliation{Department of Physics, Carnegie Mellon University, 
Pittsburgh PA\ \ 15213}

\date{\today}

\begin{abstract}
We renormalize the divergences in the energy-momentum tensor of a scalar field that begins its evolution in an effective initial state.  The effective initial state is a formalism that encodes the signatures of new physics in the structure of the quantum state of a field; in an inflationary setting, these signatures could include trans-Planckian effects.  We treat both the scalar field and gravity equivalently, considering each as a small quantum fluctuation about a spatially independent background.  The classical gravitational equations of motion then arise as a tadpole condition on the graviton.  The contribution of the scalar field to these equations contains divergences associated with the structure of the effective state.  However, these divergences occur only at the initial time, where the state was defined, and they accompany terms depending solely upon the classical gravitational background.  We define the renormalization prescription that adds the appropriate counterterms at the initial-time boundary to cancel these divergences, and illustrate it with several examples evaluated at one-loop order.
\end{abstract}


\maketitle

\section{Introduction}
\label{intro}

A distinctive element of inflation is its proposal that the structures of the universe had their birth through a purely quantum mechanism.  This mechanism relies on two essential components, the natural variations, or fluctuations, which occur for any quantum field and an extremely rapid expansion during an early stage of the universe.  It is this expansion that stretches the tiny quantum fluctuations until they are of a vast size, forming the pattern of primordial perturbations frozen into the otherwise smooth background of space-time.

Because inflation relies upon such a simple and elegant set of ingredients, its basic predictions are shared across many of its particular implementations \cite{textbooks}.  The typical model for inflation uses a quantum scalar field, called the inflaton, both to drive the expansion through its classical potential and to provide a source for the quantum fluctuations when combined with the scalar component of the gravitational fluctuations.  For this potential energy to be able to dominate the dynamics, the kinetic energy of the inflaton should be small, which in turn implies some rather stringent constraints on the shape of the potential.  The potential must be remarkably flat during the period when inflation occurs which requires that the mass and the couplings of the quantum part should be small.  As a consequence, the primordial set of perturbations produced by inflation describes a nearly featureless pattern of Gaussian noise.  Even more distinctively, inflation also predicts that this pattern should extend beyond the causal horizon that would have been expected of a universe that had not undergone an inflationary phase.  

The radiation and matter environments which come eventually to dominate the evolution of the universe after inflation ends react to these primordial perturbations with a characteristic pattern \cite{acoustic}.  In fact, exactly this pattern has been observed recently both in the cosmic microwave background \cite{wmap} as well as the structure of the universe on scales sufficiently large to be still unaffected by nonlinearities of gravitational collapse \cite{sdss}.  Moreover, the anticorrelation of the fluctuations in the polarization and in the temperature of the microwave background radiation at large scales, which---without inflation---would have been out of causal contact until only recently, is also consistent with the expectations of inflation \cite{polar}.

Despite this consistency between its predictions and what is observed, inflation nevertheless contains several unappealing properties \cite{troubles}.  The most significant of these follows directly from those two basic elements of inflation:  its quantum component and its space-time expansion.  Consider a fluctuation of a particular size at the end of inflation, which we choose to be one that will later produce some feature in the microwave background.  As we follow its evolution backwards in time, its size shrinks until it is within the inflationary Hubble horizon.  If we continue still further back, provided inflation lasted sufficiently long, this fluctuation would eventually be smaller than the Planck length, where we no longer have a perturbative description of the quantum theory.  This observation---that we might have to know all the subtleties at the Planck scale if we are to predict the patterns of structure seen in the universe---constitutes the {\it trans-Planckian problem\/} of inflation \cite{brandenberger,greens,dSgreens,emtensor,schalm,schalm2,ekp,emil,kaloper,cliff,gary,transplanck,dan}.

The trans-Planckian problem actually has two sides, one more practical and the other aesthetic.  At the practical level, all of the features of the observed universe can be explained without extending any of scales of the primordial perturbations so far back that they were smaller than the Planck scale.  If we are content to start our theory when these perturbations are larger than the Planck scale but still within the Hubble horizon, then all of their subsequent evolution can be described perturbatively.  In proceeding thus, we must make some choice about the inflaton's state when we begin the evolution.  So there is nothing that invalidates our using inflation to generate the primordial perturbations; it simply becomes must more doubtful to regard it as a ``final'' theory, that is, one which automatically chooses everything---the correct state as well---for us.  This choice corresponds to the second, aesthetic, side of the trans-Planckian problem.  To explain why a particular state is appropriate requires inevitably looking further back during inflation, or appealing to some additional mechanism, and to do so requires that we make some assumption about either what states are natural, even speaking of their features on arbitrarily small scales, or what is the appropriate description of quantum gravity---assuming that it can provide a perturbative, controlled framework.

There is, however, another approach to this problem which is much more in keeping with the principles of effective field theory \cite{eft}.  From its perspective, inflation is no longer seen as a final theory, one that can be extended to arbitrarily early times; instead, we gain a controlled method for describing the generic imprint of new physics beyond the energy scales of inflation.  An effective field theory divides phenomena according to whether they are of lower or higher energy compared with an energy or mass scale $M$, associated with some physics that does not appear directly at the lower energies.  As a field theory, the observed degrees of freedom and their symmetries determine the renormalizable operators included, but the nonrenormalizable operators, which are the artifacts of that hidden physics, are left completely general, as long as they do not violate any low energy symmetries.  What makes this approach predictive is that all of these higher order operators are accompanied by inverse powers of the scale $M$ so that in a measurement made at an energy $E<M$, for example, the effect of a particular operator associated with the new physics is suppressed by a power of $E/M$.  For a fixed experimental accuracy, only a small number of these operators will be measurable.  Thus any practical measurement requires only a finite set of unknown parameters, at least until we are able to probe the new physics directly, $E\sim M$.

Effective field theory strictly applies to the evolution of the state, and by itself does not capture how new physics may be imprinted in a particular initial state.  To do so, we introduce an effective description of the state \cite{greens,dSgreens,emtensor,schalm,schalm2,ekp,emil}.  This effective description applies a similar principle, dividing the structures in a state according to whether they vary significantly or not over a length scale $1/M$.  Here, the mass scale $M$ can be any value above the Hubble scale during inflation, $H$, which plays the same role as the measurement scale $E$ above. Furthermore, while $M$ can certainly correspond to the Planck mass, $M_{\rm pl}$, it can also more generally be any scale beyond $H$, such as the binding energy of a composite inflaton or the excitation energy of some hidden field, for example.  In an analogy with an effective field theory, the behavior of the state at lengths much longer than $1/M$ is chosen phenomenologically, in this case to match the symmetries of the coarse features of the primordial perturbations extracted from the microwave background.  At short distances, the structure of the state is parameterized by a set of moments scaling as increasing powers of $1/M$.  We shall refer to this last set of structures at the ``trans-Planckian'' part of the state, even when $M < M_{\rm pl}$.

An effective theory relies on the smallness of the ratio of the dynamically accessible scale---here that provided by inflation $H$---and the scale of new phenomena, $M$, to provide a perturbative framework.  For this perturbation theory to be sensible, we need to show how to treat processes which are sensitive to the short-distance structures.  For the purely scalar part of the theory, such processes correspond to loop corrections, which sum over all intermediate momenta and are thus also sum over the arbitrarily short-distance features of the state.  In \cite{greens,dSgreens}, we showed that such processes introduce new divergences which are confined to the initial time at which the state is defined; consequently, they are cancelled by including the appropriate operators localized on this same space-like surface.  Through this boundary renormalization, the loop corrections are rendered completely finite and small.

On the gravitational side, the expectation value of the energy-momentum tensor for the quantum part of the scalar field similarly sums together the contributions over all distance scales.  In this case, the short-distance structure of the effective state of the scalar field again produces divergences at the initial-time boundary, but these divergences are now wholly proportional to terms that depend on the classical component of gravity.  Such divergences therefore require that we also apply our boundary renormalization prescription to gravity. This program was begun in \cite{emtensor} where we treated gravity classically.  However, since the boundary divergences arise from quantum effects, it is more appropriate to treat both gravity and the scalar field on an equal footing, which is done here.  In this article, we describe the origin and the renormalization of the new boundary divergences in the energy and momentum associated with the effective initial state when we include a small quantum fluctuation in the space-time metric about a classically expanding background.

The next section reviews the behavior of gravity in the small fluctuation limit and derives the relevant parts of the action that are required for the renormalization of the energy-momentum tensor in this limit.  Sections~\ref{propagate} and \ref{evolve} summarize the effective state idea, describing first how the information in the effective state modifies the propagator and then how the time-evolution of matrix elements is defined.  Section~\ref{graviton} provides an overview of the general method for renormalizing the boundary divergences in the energy-momentum tensor which is then exemplified in the following section through three simple specific cases.  This article is part of a series \cite{greens,dSgreens,emtensor} that establishes the renormalizability and consistency of the effective description of an initial state.  Several of the applications of this formalism are outlined in the concluding section.

\section{Small fluctuations}
\label{small}

We begin with the action for gravity and the scalar field, 
\begin{equation}
S_{\cal M} = S_G + S_\Phi . 
\label{totalaction}
\end{equation}
Here we have explicitly referred to the full space-time manifold, ${\cal M}$, since later we shall also include a component for the action at the initial boundary.  Some of the terms in the boundary action are fixed by the bulk action, but others depend on the state of the system.  We write the purely gravitational part as an effective action, given by an expansion in powers of derivatives of the metric, 
\begin{equation}
S_G = \int d^4x\, \sqrt{-g}\, \bigl[ 
2\Lambda + M^2_{\rm pl} R + \alpha R^2 + \cdots \bigr] . 
\label{Gaction}
\end{equation}
The first two terms correspond to the cosmological constant contribution and to the standard Einstein-Hilbert term.  We shall study here a conformally flat space-time, so among the three available invariant terms with four derivatives, $R^2$, $R_{\mu\nu} R^{\mu\nu}$ and $R_{\lambda\mu\nu\rho} R^{\lambda\mu\nu\rho}$, we only require a linear combination of these invariants that is orthogonal to both the Gauss-Bonnet term, 
\begin{equation}
R^2 - 4 R_{\mu\nu} R^{\mu\nu} + R_{\lambda\mu\nu\rho} R^{\lambda\mu\nu\rho} , 
\label{gaussbonnet}
\end{equation}
which being a total derivative does not affect the dynamics of gravity, and the squared Weyl term, 
\begin{equation}
C_{\lambda\mu\nu\rho} C^{\lambda\mu\nu\rho} 
= {\textstyle{1\over 3}} R^2 - 2 R_{\mu\nu} R^{\mu\nu} 
+ R_{\lambda\mu\nu\rho} R^{\lambda\mu\nu\rho} , 
\label{weylweyl}
\end{equation}
which vanishes for conformally flat metrics.  The simplest such orthogonal combination is provided by the $R^2$ term alone.  A completely general set of fluctuations about the conformally flat background will generate the squared Weyl term, but since we shall confine the calculation to how the quantum parts of the scalar field affect the classical part of the geometry, the $R^2$ term is by itself sufficient for this purpose.

The universe at large scales or early times appears homogeneous and isotropic and a variety of measurements are consistent with it being spatially flat as well; by appropriately choosing the time coordinate, such a background geometry can be written in a conformally flat form, 
\begin{equation}
ds^2 = g_{\mu\nu}\, dx^\mu dx^\nu 
= a^2(\eta) \bigl[ d\eta^2 - d\vec x\cdot d\vec x \bigr] + \cdots . 
\label{background}
\end{equation}
The evolution of the classical background is thus described by a single function, the scale factor $a(\eta)$, and the fractional rate by which it changes corresponds to the Hubble scale, 
\begin{equation}
H(\eta) = {a'\over a^2} \equiv {1\over a^2} {\partial a\over\partial\eta} , 
\label{hubble}
\end{equation}
which sets the natural dynamical scale associated with the background.  In the inflationary picture, the original set of perturbations arise from quantum fluctuations about this background so we include a small quantum piece in the metric, 
\begin{equation}
g_{\mu\nu} 
= a^2(\eta)\, \bigl[ \eta_{\mu\nu} + h_{\mu\nu}(\eta,\vec x) \bigr] , 
\label{smallhdef}
\end{equation}
where $\eta_{\mu\nu}$ is the Minkowski metric.  Note that we shall always raise and lower the indices of $h_{\mu\nu}$ with $\eta_{\mu\nu}$

The source for the expansion of the space-time is here provided by the scalar field, $\Phi(\eta,\vec x)$; for simplicity we assume that this field does not have any self-interactions and that it only interacts minimally with gravity, 
\begin{equation} 
S_\Phi = \int d^4x\, \sqrt{-g}\, \bigl[ 
{\textstyle{1\over 2}} g^{\mu\nu} \partial_\mu\Phi \partial_\nu\Phi 
- {\textstyle{1\over 2}} m^2 \Phi^2 \bigr] . 
\label{Phiaction}
\end{equation}
We divide this field as well into separate pieces for the classical zero mode, $\phi(\eta)$, and the quantum fluctuations, $\varphi(\eta,\vec x)$, so that 
\begin{equation}
\Phi(\eta,\vec x) = \phi(\eta) + \varphi(\eta, \vec x) . 
\label{fielddef}
\end{equation}

The classical equations of motion for the background, $\{ a(\eta), \phi(\eta) \}$, arise as renormalization conditions imposed on the fluctuations.  More precisely, the vanishing of the tadpole graphs for the scalar and the metric fluctuations,
\begin{equation}
\langle 0_{\rm eff}(\eta) | \varphi(\eta,\vec x) | 0_{\rm eff}(\eta) \rangle 
= 0 , 
\label{phitadpole}
\end{equation}
and 
\begin{equation}
\langle 0_{\rm eff}(\eta) | h_{\mu\nu}(\eta,\vec x) | 0_{\rm eff}(\eta) \rangle = 0 , 
\label{tadpole}
\end{equation}
correspondingly yield the Klein-Gordon equation for $\phi(t)$ and the gravitational equations of motion, along with their higher order loop corrections.  The advantage of this approach, compared with treating gravity only classically, is that it is more straightforward to understand the renormalization of the gravitational equation, particularly that component of the renormalization that is associated with the boundary divergences produced by the initial state, since these matrix elements contain an additional time integral associated with evolving the state from its initial form defined at $\eta_0$ to an arbitrary later time $\eta$.  In these tadpole conditions, we have referred to the effective state of the system, $|0_{\rm eff}(\eta)\rangle$.  How it is defined initially and how it evolves will be the subject of the next two sections.

In this language, the scalar field requires the renormalization of the parameters in the gravitational action, $\{ \Lambda, M_{\rm pl}, \alpha \}$, through a one-loop correction to the graviton tadpole condition.  To this order, it is sufficient to extract only the pieces of the action that are linear in $h_{\mu\nu}$.  For the gravitational part of the action, we thus only need 
\begin{equation}
\sqrt{-g} = a^4 \bigl[ 1 + {\textstyle{1\over 2}} h 
+ {\cal O}\bigl( h^2 \bigr) \bigr] 
\label{detgapprox}
\end{equation}
and 
\begin{eqnarray}
R &\!\!\!=\!\!\!& 
{6\over a^3} a^{\prime\prime} 
- {6\over a^3} h^{00} a^{\prime\prime} 
- {3\over a^3} a' \bigr[ 
2 \partial_\mu h^{\mu 0} - \eta^{\mu 0}\partial_\mu h 
\bigr]
\nonumber \\
&& 
- {1\over a^2} \bigr[ 
\partial_\mu \partial_\nu h^{\mu\nu} 
- \eta^{\mu\nu}\partial_\mu \partial_\nu h 
\bigr]
+ {\cal O}\bigl( h^2 \bigr) , 
\label{Rapprox}
\end{eqnarray}
where 
\begin{equation} 
h = \eta_{\mu\nu} h^{\mu\nu} = h^{00} - \delta_{ij} h^{ij} 
\label{trh}
\end{equation}
is the trace of the fluctuations.  Denoting the gravitational Lagrange density by ${\cal L}_G$, to linear order we have
\begin{eqnarray}
{\cal L}_G &\!\!\!=\!\!\!& 
\sqrt{-g}\, \bigl[ 2\Lambda + M_{\rm pl}^2 R + \alpha R^2 \bigr] 
\nonumber \\
&\!\!\!=\!\!\!& 
{\cal L}_G^0 + a^2 h^{\mu\nu} \tilde G_{\mu\nu} 
+ D^{(1)}_G + D^{(2)}_G + {\cal O}\bigl( h^2 \bigr) . 
\label{LGlinear}
\end{eqnarray}
The first term, 
\begin{equation}
{\cal L}_G^0 = 2 \Lambda a^4 - 6 M_{\rm pl}^2 a a^{\prime\prime} 
+ 36 \alpha \biggl( {a^{\prime\prime}\over a} \biggr)^2 , 
\label{LGzero}
\end{equation}
does not affect the dynamics of the fluctuating part of the fields.  The tensor $\tilde G_{\mu\nu}$ is essentially the standard Einstein tensor supplemented by the cosmological constant term as well as the four-derivative terms.  For an isotropically expanding space-time, its components are 
\begin{eqnarray}
\tilde G_{00} &\!\!\!=\!\!\!& 
\Lambda a^2 
+ 3 M^2_{\rm pl} \biggl( {a'\over a} \biggr)^2 
\nonumber \\
&& 
+ {18\alpha\over a^2} \biggl[ 
2 {a^{\prime\prime\prime}\over a} {a'\over a} 
- \biggl( {a^{\prime\prime}\over a} \biggr)^2 
- 4 {a^{\prime\prime}\over a} \biggl( {a'\over a} \biggr)^2 
\biggr] 
\nonumber \\
\tilde G_{0i} &\!\!\!=\!\!\!& 0
\nonumber \\
\tilde G_{ij} &\!\!\!=\!\!\!& 
\delta_{ij}\, \biggl\{ 
- \Lambda a^2 
- M^2_{\rm pl} \biggl[ 
2 {a^{\prime\prime}\over a} - \biggl( {a'\over a} \biggr)^2 
\biggr] 
\nonumber \\
&& 
- {12\alpha\over a^2} \biggl[ 
{a^{\prime\prime\prime\prime}\over a} 
- 5 {a^{\prime\prime\prime}\over a} {a'\over a} 
- {5\over 2} \biggl( {a^{\prime\prime}\over a} \biggr)^2 
+ 8 {a^{\prime\prime}\over a} \biggl( {a'\over a} \biggr)^2 
\biggr] \biggr\} . 
\nonumber \\
\label{Gtildedef}
\end{eqnarray}
The remaining two terms in the gravitational Lagrange density are total derivatives, 
\begin{eqnarray} 
D_G^{(1)} &\!\!\!=\!\!\!&
- M_{\rm pl}^2 \partial_\mu \bigl\{ 
4 aa' h^{\mu 0} - aa' \eta^{\mu 0} h 
\nonumber \\
&&\qquad\qquad
+ a^2 \partial_\nu h^{\mu\nu} 
- a^2 \eta^{\mu\nu} \partial_\nu h \bigr\}
\nonumber \\
D_G^{(2)} &\!\!\!=\!\!\!&
12\alpha \partial_\mu \biggl\{ 
- {a^{\prime\prime}\over a} 
\bigl[ \partial_\nu h^{\mu\nu} - \eta^{\mu\nu} \partial_\nu h \bigr]
\nonumber \\
&&\qquad\quad
+ \biggl[ {a^{\prime\prime\prime}\over a} 
- 7 {a^{\prime\prime}\over a} {a'\over a} \biggr] h^{\mu 0} 
\nonumber \\
&&\qquad\quad
- \biggl[ {a^{\prime\prime\prime}\over a} 
- 4 {a^{\prime\prime}\over a} {a'\over a} \biggr] \eta^{\mu 0} h 
\biggr\} . 
\label{LGdervs}
\end{eqnarray}

These derivative terms are the first instances of where we shall require the addition of an action at the initial boundary, although unlike those we shall encounter later, they are not related to the initial state chosen.  These terms arose as we moved derivatives from the fluctuating part of the field, $h_{\mu\nu}$, to the background.  Since we do not assume that there is any spatial boundary to the space-time, or alternatively that the fields diminish sufficiently quickly if it does, the spatial derivatives in Eq.~(\ref{LGdervs}) can be neglected.  What remains is a contribution at the initial boundary at $\eta=\eta_0$ which still contains conformal time derivatives---normal derivatives---of the fields, 
\begin{eqnarray}
&&\!\!\!\!\!\!
\int_{\eta_0}^0 d\eta \int d^3\vec x\, 
\bigl\{ D_G^{(1)} + D_G^{(2)} \bigr\}
\label{LGdervsBA} \\
&\!\!\!=\!\!\!&
M_{\rm pl}^2 \int_{\eta_0} d^3\vec x\, 
\bigl\{ a^2 \partial_0 h^{00} - a^2 \partial_0 h 
+ 4 aa' h^{00} - aa' h \bigr\}
\nonumber \\
&&
+ 12\alpha \int_{\eta_0} d^3\vec x\, 
\biggl\{ 
{a^{\prime\prime}\over a} 
\bigl[ \partial_0 h^{00} - \partial_0 h \bigr]
\nonumber \\
&&\qquad\quad
- \biggl[ {a^{\prime\prime\prime}\over a} 
- 7 {a^{\prime\prime}\over a} {a'\over a} \biggr] h^{00} 
+ \biggl[ {a^{\prime\prime\prime}\over a} 
- 4 {a^{\prime\prime}\over a} {a'\over a} \biggr] h 
\biggr\} . 
\nonumber
\end{eqnarray}
The normal derivatives within the first terms are cancelled by adding a Gibbons-Hawking action \cite{gh} at the initial boundary, 
\begin{equation}
S_{\rm GH} = 2 M_{\rm pl}^2 \int_{\eta_0} d^3\vec x\, \sqrt{\hat g} K .
\label{GHaction}
\end{equation}
Here $\hat g$ corresponds to the determinant of the induced metric on the initial surface.  If we establish a time-like unit normal vector, $n_\mu$, to this surface, then the induced metric is obtained from the full metric through
\begin{equation}
\hat g_{\mu\nu} = g_{\mu\nu} - n_\mu n_\nu .
\label{induced}
\end{equation}
$K$ is the trace of the extrinsic curvature, $K_{\mu\nu}$, defined by the projection of the covariant derivative of the normal, 
\begin{equation}
K_{\mu\nu} = \hat g_\mu^{\ \lambda} \nabla_\lambda n_\nu , 
\label{extrinsic}
\end{equation}
and to first order in the small fluctuations this trace is 
\begin{eqnarray}
K &\!\!\!=\!\!\!&
3 {a'\over a^2} - 3 {a'\over a^2} h^{00} 
- {1\over 2} {1\over a} \partial_0 h^{00} 
+ {1\over 2} {1\over a} \partial_0 h 
\nonumber \\
&& 
+ \partial_i \biggl[ {1\over a} \delta^{ij} h_{0j} \biggr] 
+ {\cal O}\bigl( h^2 \bigr) . 
\label{Korderh}
\end{eqnarray}
Thus to this linear order, the Gibbons-Hawking term has precisely the correct form to cancel the normal derivatives that resulted from the $D_G^{(1)}$ term of the bulk Lagrangian, 
\begin{eqnarray}
S_{\rm GH} &\!\!\!=\!\!\!& 
M_{\rm pl}^2 \int_{\eta_0} d^3\vec x\, 
\bigl\{ 
6 aa' - a^2 \partial_0 h^{00} + a^2 \partial_0 h 
\nonumber \\
&&\qquad
- 3 aa' h^{00} - 3 aa' h 
+ \partial_i \bigl[ 2 a^2 \delta^{ij} h_{0j} \bigr] 
+ \cdots \bigr\} .
\nonumber \\
&&\label{GHacttoh}
\end{eqnarray}

The normal derivatives that resulted from integrating the $R^2$ terms by parts can similarly be removed by including the appropriate surface terms.  Hereafter we shall assume that these boundary actions, the Gibbons-Hawking action and its higher derivative generalization, have been included in the theory so that we can focus entirely upon that part of the boundary action that is needed to remove the divergences associated with the effective state.

The second component of the theory, the scalar field, couples to $h_{\mu\nu}$ at linear order through its energy-momentum tensor.  Since this field itself contains both a classical zero mode and a quantum fluctuation, it is useful to separate this tensor into a piece for the classical part, 
\begin{equation}
T^{\rm cl}_{\mu\nu} = \partial_\mu\phi \partial_\nu\phi 
- {\textstyle{1\over 2}} \eta_{\mu\nu}\partial_\lambda\phi\partial^\lambda\phi 
+ {\textstyle{1\over 2}} \eta_{\mu\nu} a^2m^2 \phi^2 , 
\label{clemt}
\end{equation}
and a term for the energy-momentum of the fluctuating part of the field, 
\begin{equation}
\hat T_{\mu\nu} = \partial_\mu\varphi \partial_\nu\varphi 
- {\textstyle{1\over 2}} \eta_{\mu\nu} 
\partial_\lambda\varphi \partial^\lambda\varphi 
+ {\textstyle{1\over 2}} \eta_{\mu\nu} a^2m^2 \varphi^2 . 
\label{emt}
\end{equation}
Note that unlike the classical part, $\hat T_{\mu\nu}$ is an operator.  Isolating the terms linear in $h_{\mu\nu}$, the scalar field's Lagrangian then yields, 
\begin{eqnarray}
{\cal L}_\Phi &\!\!\!=\!\!\!& 
\sqrt{-g}\, \bigl[ 
{\textstyle{1\over 2}} g^{\mu\nu} \partial_\mu\Phi \partial_\nu\Phi 
- {\textstyle{1\over 2}} m^2 \Phi^2 \bigr] 
\nonumber \\
&\!\!\!=\!\!\!& 
\cdots 
- {\textstyle{1\over 2}} a^2 h^{\mu\nu} \bigl\{
T^{\rm cl}_{\mu\nu} + \hat T_{\mu\nu} 
+ \partial_\mu\phi\partial_\nu\varphi + \partial_\mu\varphi \partial_\nu\phi \nonumber \\ 
&&\qquad 
- \eta_{\mu\nu} \partial_\lambda\phi \partial^\lambda\varphi 
+ \eta_{\mu\nu} a^2m^2 \phi\varphi \bigr\} 
+ {\cal O} \bigl( h^2 \bigr) . \qquad
\label{Lphilinear}
\end{eqnarray} 
We have not defined a separate tensor for the cross-term containing both $\phi$ and $\varphi$ since to the order to which we shall evaluate the graviton tadpole, Eq.~(\ref{tadpole}), the scalar tadpole condition in Eq.~(\ref{phitadpole}) requires that the leading contribution of this cross-term vanishes.

Before concluding this section we gather together the necessary parts of the ``bare'' action for evaluating the graviton tadpole to one-loop order in the scalar field, 
\begin{equation}
S_{\cal M} = S_h^0 + S_\varphi^0 + S_I + \cdots . 
\label{reorderact}
\end{equation}
The first two terms are those quadratic in the quantum parts of the scalar field, 
\begin{equation}
S_\varphi^0 = \int_{\eta_0}^0 d\eta\, a^2 \int d^3\vec x\, 
\Bigl\{ {\textstyle{1\over 2}} \eta^{\mu\nu} \partial_\mu\varphi \partial_\nu\varphi - {\textstyle{1\over 2}} a^2m^2 \varphi^2 
\Bigr\}  
\label{freephi}
\end{equation}
and the metric; they define the free propagators for the scalar field and graviton.  Since the graviton only appears as an external leg, common to all of the leading contributions to the tadpole in Eq.~(\ref{tadpole}), we shall not need the detailed form of its propagator, so we have not explicitly written $S_h^0$.  The final term contains all the bulk terms that are linear in $h_{\mu\nu}$, 
\begin{equation}
S_I = \int_{\eta_0}^0 d\eta\, \int d^3\vec x\, 
{\textstyle{1\over 2}} a^2 h^{\mu\nu} 
\bigl\{ 2\tilde G_{\mu\nu} - T_{\mu\nu}^{\rm cl} - \hat T_{\mu\nu} \bigr\} , 
\label{interaction}
\end{equation}
and which corresponds to only a small component of the full set of gravitational interactions, but which is the part responsible for the leading corrections to the tadpole.

\section{Propagation}
\label{propagate}

The particular element we wish to explore in this article is the role and the renormalizability of the trans-Planckian parts of the initial state and although we shall be calculating the graviton tadpole, it is the effect of the state of the scalar field that we shall need to describe in detail, since it provides the leading quantum correction to the tadpole through the scalar loop.  In this section therefore we review how the choice of the effective state influences the propagator of the scalar field.  Much of this material is developed more fully in \cite{greens,dSgreens,emtensor}, so the presentation here will be brief, although still reasonably self-contained.

The usual method for selecting the state, generalizing the approach applied for the $S$-matrix description of flat space, is to restrict first to the maximally symmetric solutions of the free Klein-Gordon equation, 
\begin{equation}
\varphi_k^{\prime\prime} + 2aH \varphi'_k 
+ \bigl( k^2 + a^2 m^2 \bigr) \varphi_k = 0 , 
\label{KG}
\end{equation}
and then to choose that unique state among these that satisfies a condition on its asymptotic short-distance behavior.  Here the $\varphi_k(\eta)$ are the Fourier modes of the field, which are the coefficients of an operator expansion of the field, 
\begin{equation}
\varphi(\eta,\vec x) = \int {d^3\vec k\over (2\pi)^3}\, 
\bigl[ \varphi_k(\eta) e^{i\vec k\cdot\vec x} a_{\vec k}
+ \varphi_k^*(\eta) e^{-i\vec k\cdot\vec x} a_{\vec k}^\dagger \bigr] . 
\label{opexpand}
\end{equation}

The standard, or Bunch-Davies \cite{bunch}, vacuum corresponds to the state that matches with the flat space vacuum state at distances much smaller than those at which the curvature of the background is much apparent, $k\gg H$.  Depending upon the model, during inflation the Hubble scale can be as large as a few orders of magnitude below the Planck scale, $M_{\rm pl}$, so that the state is being chosen by appealing to its behavior in a regime close to that in which a perturbative description of gravity breaks down.  The standard vacuum also does not allow for structures that appear at some intermediate scale, $H\ll M \ll M_{\rm pl}$, and that differ substantially over this range from the Bunch-Davies vacuum even if well beyond that scale, $k\gg M$, the state again approaches the usual vacuum.  Such effects can arise, for example, when the scalar is coupled to another, excited field whose excitations occur with an energy $M$ \cite{cliff}.

To be able to model such effects as these, we shall allow more general structures in the state by defining it {\it effectively\/} at an initial time $\eta=\eta_0$.  The idea behind this description is that features can be characterized as either long or short distance structures relative to a scale of new physics $1/M$.  This scale can be either dynamical, as would be the case for example of a composite inflaton or an inflaton coupled to a second, excited field, or ``fundamental'' in its origin.  The latter case covers a broad class of effects depending upon the particular model for physics near and above the Planck scale, and can include nonstandard short-distance structures such as those related to a breakdown of the usual uncertainty relation \cite{gary} or some form of space-time non-commutativity \cite{smolin}, among many other possibilities.

In an effective field theory, which is used to describe how unknown physics can affect the {\it evolution\/} of a system, the long-distance properties are fixed to agree with what is observed.  A similar principle applies to the effective state.  In the case of inflation, the natural choice for this range of scales is the Bunch-Davies vacuum, since this state produces the nearly flat primordial power spectrum that has been extracted from measurements of the cosmic background radiation.  If we denote the modes associated with this state by $U_k(\eta)$, then a general solution to Eq.~(\ref{KG}) can be expressed in terms of these modes by 
\begin{equation}
\varphi_k(\eta) = {U_k(\eta) + f_k\, U_k^*(\eta)\over\sqrt{1-f_kf_k^*}} , 
\label{genmodes}
\end{equation}
where we have fixed the normalization of the state by requiring the field to satisfy the equal-time commutation relation, 
\begin{equation}
\bigl[ \pi(\eta,\vec x), \varphi(\eta,\vec y) \bigr] 
= - i\, \delta^3(\vec x-\vec y) 
\qquad 
\pi = a^2\, \varphi' . 
\label{etcr}
\end{equation}
The vacuum state associated with these modes will be denoted by $|0_{\rm eff}\rangle$. 

The function $f_k$ describes the structure of the initial state.  The usual requirement is that $f_k$ should decrease sufficiently fast as $k\to\infty$, typically faster than $k^{-4}$ \cite{emil}, which in effect only allows long-distance modifications of the state.  Here we are most interested in the role of the short-distance trans-Planckian details of the state, so $f_k$ will not need to satisfy this condition.  Because such states differ most substantially from the Bunch-Davies state at short-distance, processes that sum over all scales---such as loop corrections or the expectation value of the energy-momentum tensor---will contain new divergences. But because they are associated with the initial state, they are confined to the initial-time surface, $\eta=\eta_0$, and are removed by adding local counterterms there.  Here we shall show how this boundary renormalization is applied to the divergences in the expectation value of the energy-momentum tensor.

In practice we shall fix the initial state through a boundary condition on the modes applied at the initial time,
\begin{equation}
\partial_\eta \varphi_k(\eta)\bigr|_{\eta=\eta_0} = - i\varpi_k \varphi_k(\eta_0) ; 
\label{bconvarphi}
\end{equation}
the momentum-dependent boundary condition $\varpi_k$ determines the initial state structure function through 
\begin{equation}
f_k = - {\varpi_k U_k(\eta_0) - i U'_k(\eta_0)\over 
\varpi_k U_k^*(\eta_0) - i U_k^{*\prime}(\eta_0) } . 
\label{fkasombar}
\end{equation}
What is more important for the dynamics and renormalization of the theory is the prescription for how the information in the initial state propagates forward in time.  To begin, we can impose an analogous boundary condition on the momentum modes of the propagator, 
\begin{eqnarray}
\partial_\eta G_k(\eta,\eta')\bigr|_{\eta=\eta_0} 
&\!\!\!=\!\!\!& i\varpi_k^* G_k(\eta_0,\eta')
\nonumber \\
\partial_{\eta'} G_k(\eta,\eta')\bigr|_{\eta'=\eta_0} 
&\!\!\!=\!\!\!& i\varpi_k^* G_k(\eta,\eta_0)
\label{bconG}
\end{eqnarray}
where the spatial representation of the propagator is
\begin{eqnarray}
\langle 0_{\rm eff} | T \bigl( \varphi(x)\varphi(x') \bigr) | 0_{\rm eff} \rangle 
&\!\!\!=\!\!\!& -i G(x,x') 
\label{Feynman} \\
&\!\!\!=\!\!\!& 
-i \int {d^3\vec k\over (2\pi)^3}\, e^{i\vec k\cdot (\vec x-\vec x')}
G_k(\eta,\eta') . 
\nonumber 
\end{eqnarray}
Note that $T\big( \cdots \bigr)$ is not the usual time-ordering, but one consistent with Eq.~(\ref{bconG}).

By themselves, these boundary conditions are not quite sufficient to fix the propagator completely since there is a remaining freedom between choosing the modes and choosing the specific form of the time-ordering.  This remaining ambiguity is removed \cite{dSgreens} by demanding that the loop corrections should be free of pinched singularities \cite{einhorn} or nonrenormalizable divergences \cite{fate} which results in a unique choice among the possible time-orderings consistent with Eq.~(\ref{bconG}), 
\begin{eqnarray}
G_k(\eta,\eta') &\!\!\!=\!\!\!& 
\Theta(\eta-\eta')\, U_k(\eta) U_k^*(\eta') 
\label{prop} \\
&& 
+ \Theta(\eta'-\eta)\, U_k^*(\eta) U_k(\eta') 
+ f_k^* U_k(\eta) U_k(\eta') . 
\nonumber 
\end{eqnarray}

One reason that we have reviewed the origin and the structure of the propagator in some detail is that the correction to the expectation value of the energy-momentum tensor from the scalar field is in its essence a loop correction to the graviton tadpole.  From a more general perspective, the propagator properly encodes the forward evolution of the information contained in the initial state and thus provides the paradigm for how to include this same information in any other matrix element of the effective theory, including this expectation value of the energy-momentum.

The energy-momentum tensor is essentially a two-point operator upon which a classical derivative operator acts.  Since the derivatives act on each field separately, the standard method \cite{pointsplit} is to evaluate the fields at different points, $x=(\eta,\vec x)$ and $x'=(\eta',\vec x')$, extract the derivatives, take the expectation value, and then return to the limit where the points coincide, 
\begin{equation}
T_{\mu\nu} = -i \lim_{x'\to x} 
\bigl[ \hat\partial_\mu\hat\partial'_\nu 
- {\textstyle{1\over 2}} \eta_{\mu\nu} \hat\partial_\lambda \hat\partial^{\prime\lambda} 
+ {\textstyle{1\over 2}} \eta_{\mu\nu} a^2m^2 \bigr] G(x,x') . 
\label{Tmunups}
\end{equation}
Here $T_{\mu\nu}$ is the expectation value of the energy-momentum operator for the scalar field in the effective state, 
\begin{equation}
T_{\mu\nu}(\eta) \equiv 
\langle 0_{\rm eff} | \hat T_{\mu\nu}(\eta, \vec x) | 0_{\rm eff} \rangle , 
\label{expTmunu}
\end{equation}
and is therefore a classical function.

There are only two subtleties in writing the energy-momentum tensor thus, the first and simpler being that the derivatives which acted only on the fields should not act on the $\Theta$-functions associated with the time-ordering when we have commuted them with the action of taking the expectation value.  Therefore we have explicitly written the derivatives as $\hat\partial_\mu$ to indicate that the time derivatives do not act upon the $\Theta$-functions in the propagator.

The second subtlety is that in writing the expectation value of the energy-momentum tensor in terms of the propagator we have implicitly chosen a particular time-ordering of the point-split fields.  It has been made since this choice correctly includes the boundary contribution from the effective initial state---the last term in Eq.~(\ref{prop})---and so it naturally leads to a renormalizable set of boundary divergences.  Because the remaining $\Theta$-functions in the propagator are explicitly unaffected by the time-derivatives as we have defined them, the standard parts of $T_{\mu\nu}$ are unaffected by choosing other possible time-orderings.  This last point is important since the Schwinger-Keldysh formalism, which we introduce in the next section, applies different time-orderings for the standard parts of the propagator---the first two terms in Eq.~(\ref{prop}).  This insensitivity to this part of the time-orderings means that each of the four possible propagators that we shall encounter in the Schwinger-Keldysh approach all yield the same result for $T_{\mu\nu}$ when we allow the split points to coincide once again, $x'\to x$.

The expectation value of the energy-momentum tensor inherits the same classical symmetries as the background, implying that its only nonvanishing components are those lying along the diagonal and that moreover these are specified by two time-dependent functions, 
\begin{equation}
T_{00}(\eta) = a^2(\eta)\, \rho(\eta)
\qquad
T_{ij}(\eta) = a^2(\eta)\, p(\eta)\, \delta_{ij} , 
\label{rhopdef}
\end{equation}
which are interpreted as the energy density $\rho(\eta)$ contained in the scalar field configuration and the pressure $p(\eta)$ exerted by it.  The conservation of the energy and momentum contained within the field, $\nabla_\lambda T^\lambda_{\ \mu}=0$, in turn implies that these two functions are related, 
\begin{equation}
\rho' = - 3aH(\rho+p) .
\label{conserve}
\end{equation}

Evaluating $T_{\mu\nu}(\eta)$ as in Eq.~(\ref{Tmunups}), we obtain the following expressions for the energy density, 
\begin{eqnarray}
\rho &\!\!\!=\!\!\!& 
{1\over 2} {1\over a^2} \int {d^3\vec k\over (2\pi)^3}\, 
\bigl\{ U_k' U_k^{\prime *} + (k^2 + a^2m^2) U_k U_k^* 
\nonumber \\
&&\qquad\qquad 
+ f_k^* \bigl[ U_k' U_k' + (k^2 + a^2m^2) U_k U_k \bigr] \bigr\} , 
\qquad 
\label{rhoval}
\end{eqnarray}
and the pressure, 
\begin{eqnarray}
p &\!\!\!=\!\!\!& - \rho 
+ {1\over a^2} \int {d^3\vec k\over (2\pi)^3}\, 
\bigl\{ U_k' U_k^{\prime *} + {\textstyle{1\over 3}} k^2 U_k U_k^* 
\nonumber \\
&&\qquad\qquad\qquad 
+ f_k^* \bigl[ U_k' U_k' + {\textstyle{1\over 3}} k^2 U_k U_k \bigr] \bigr\} , 
\qquad 
\label{pval} 
\end{eqnarray}
written in terms of the Bunch-Davies modes.  In determining the divergence structure and renormalization of the energy-momentum, it will be useful to separate the parts produced by the effective initial state and those which occur universally, even in the standard Bunch-Davies vacuum for which a standard method of renormalization was established long ago \cite{books}.  Therefore we shall write 
\begin{equation}
\rho = \rho_{\rm bulk} + \rho_{\rm surf}, 
\qquad\quad 
p = p_{\rm bulk} + p_{\rm surf} 
\label{bulksurf}
\end{equation}
where the bulk components are simply those obtained by setting $f_k\to 0$ in Eqs.~(\ref{rhoval}) and (\ref{pval}) and the surface components are composed of the complementary terms.

\subsection{An adiabatic expansion}

In an expanding background, it is quite often not possible to solve for the exact behavior of the mode functions, even those associated with the Bunch-Davies state, $U_k(\eta)$, in terms of which we have defined the modes for the more general effective states.  However, what is relevant here---for extracting the divergences in the energy-momentum of the scalar field---is only the behavior of the modes at short distances; more of this behavior needs to be understood the worse the divergence, but in any particular case only a finite part of the full space-time dependence of $U_k(\eta)$ is needed.  This observation suggests that we should apply an approximation scheme based on diminishing powers of the spatial momentum at short distances.  This approach very similar to the adiabatic expansion \cite{adiabatic}.

We begin by writing the Bunch-Davies modes in terms of a real, generalized frequency function, $\Omega_k(\eta)$, through
\begin{equation}
U_k(\eta) = {e^{-i\int_{\eta_0}^\eta d\eta\, \Omega_k(\eta')}\over a(\eta) \sqrt{2\Omega_k(\eta)}} . 
\label{BDasOm}
\end{equation}
Substituting this expression into the Klein-Gordon equation yields a differential equation for this function, 
\begin{equation}
\Omega_k^2 = k^2 + a^2 m^2 - {a^{\prime\prime}\over a} 
- {1\over 2} {\Omega^{\prime\prime}\over\Omega_k} 
+ {3\over 4} \biggl( {\Omega'\over\Omega_k} \biggr)^2 . 
\label{KGomega}
\end{equation}
The adiabatic expansion assumes that the derivative of a term in the expansion is small in comparison any term of equal or lower order, so that the $(n+1)^{\rm st}$ term in the expansion is determined by the derivative of the $n^{\rm th}$ term.  Therefore, if we let 
\begin{equation}
\Omega_k = \Omega_k^{(0)} + \Omega_k^{(1)} + \cdots 
+ \Omega_k^{(n)} + \cdots , 
\label{Omegaasns}
\end{equation}
and we start the series by defining the $0^{\rm th}$ order term to be
\begin{equation}
\Omega_k^{(0)} = \sqrt{ \omega_k^2 - {a^{\prime\prime}\over a} } , 
\label{zeroth}
\end{equation}
then the next term is given by 
\begin{equation}
\Omega_k^{(1)} = 
- {1\over 4} {\Omega^{(0)\prime\prime}\over(\Omega_k^{(0)})^2} 
+ {3\over 8} {(\Omega^{(0)\prime})^2\over(\Omega_k^{(0)})^3} . 
\label{first}
\end{equation}
In the $0^{\rm th}$ order term we have introduced the frequency 
\begin{equation}
\omega_k^2 = k^2 + a^2 m^2 , 
\label{omegadef}
\end{equation}
which very closely resembles the usual flat-space frequency.  Because $\omega_k$ at short distances essentially becomes $k$ and yet, due to the mass term, still remains finite thereby avoiding infrared divergences at large distances, we shall use $\omega_k$ as our basic expansion parameter.

Although the adiabatic expansion is not in its essence a short-distance expansion, it is straightforward enough to express the terms in the adiabatic approximation in powers of $\omega_k$, truncated to the needed order.  For example, later we shall need to expand $\Omega_k$ to order $\omega_k^{-3}$ to treat the divergences in the expectation value of the energy-momentum tensor that are independent of the state we have chosen.  To do so, we only require the first adiabatic correction, Eq.~(\ref{first}), since the second correction is already of order $\omega_k^{-5}$, 
\begin{eqnarray}
\Omega_k &\!\!\!=\!\!\!& 
\omega_k - {1\over 2} {1\over\omega_k} {a^{\prime\prime}\over a} 
- {1\over 4} {a^2m^2\over\omega_k^3} \biggl[ 
{a^{\prime\prime}\over a} - \biggl( {a'\over a} \biggr) \biggr] 
\nonumber \\
&& 
+ {1\over 8} {1\over\omega_k^3} \biggl[ 
{a^{\prime\prime\prime\prime}\over a} 
- 2 {a^{\prime\prime\prime}\over a} {a'\over a} 
- 2 \biggl( {a^{\prime\prime}\over a} \biggr)  
+ 2 {a^{\prime\prime}\over a} \biggl( {a'\over a} \biggr) \biggr] 
\nonumber \\
&& 
+ {\cal O} \biggl( {1\over\omega_k^5} \biggr) . 
\label{Omegato5}
\end{eqnarray}

Renormalizing a completely general effective state would require---in principle---evaluating the adiabatic expansion to an arbitrary order.  However, since the higher order details of the effective state are suppressed by more powers of $H/M$, for a particular aspect of the state only a finite number of adiabatic corrections are needed.  In the case of the leading trans-Planckian signal, we must expand $\Omega_k$ to order $\omega_k^{-4}$, for which the first adiabatic correction is still sufficient.

\subsection{The structure function of the effective state}

Until now we have not provided an explicit form for the structure of the initial state, which is described by the momentum-dependent function $f_k$.  We therefore conclude this section by stating a fairly general parameterization of this structure function based upon $\omega_k$ which we just introduced in Eq.~(\ref{omegadef}).  This frequency has the aforementioned advantage of being finite at long distances so that we can write a power series expansion for $f_k$ without introducing any unnecessary infrared divergences, 
\begin{equation}
f_k = \sum_{n=1}^\infty c_n 
\biggl( {a(\eta_0)m\over\omega_k(\eta_0)} \biggr)^n
+ \sum_{n=0}^\infty d_n 
\biggl( {\omega_k(\eta_0)\over a(\eta_0)M} \biggr)^n . 
\label{fkdef}
\end{equation}
The two sums correspond respectively to the long and short distance features of the state, and the terms within the second series describe the ``trans-Planckian'' features of the state, although as mentioned earlier the scale $M$ does not need necessarily to equal the Planck mass.  Although the first sum is of diminishing importance in the short-distance regime, where $\omega_k\sim k\to\infty$, its first few terms can nevertheless produce boundary divergences.  This behavior is very analogous to the standard renormalization of classical gravity in the presence of a quantum field which also requires the renormalization of the long-distance properties of the gravitational theory---the cosmological constant in particular---as a result of summing over the modes of the scalar field at short-distances.  For a few of the terms in the first sum, their resulting divergences will correspondingly require relevant counterterms at the boundary.

\section{Evolution}
\label{evolve}

The continuous expansion of the background, where scales shrink further and further the farther back we extend inflation, combined with our ignorance about the specific details of the early inflationary epoch and what might have preceded it together require that we should use a formalism that never evaluates any quantity prior to the initial time when the effective theory remains valid.  In such a setting, the standard over-all space-time approach of the $S$-matrix is no longer appropriate.  Instead, we define the state at an initial time, $|0_{\rm eff}(\eta_0)\rangle$, and then evolve it to a later time, $|0_{\rm eff}(\eta)\rangle$, where we use it to evaluate the expectation value of some operator, ${\cal O}(\eta)$, 
\begin{equation}
\langle 0_{\rm eff}(\eta) | {\cal O}(\eta) | 0_{\rm eff}(\eta) \rangle . 
\label{expeg}
\end{equation}
Later, we shall let ${\cal O}(\eta)$ be a single graviton so that this expectation value becomes the graviton tadpole given in Eq.~(\ref{tadpole}).  The approach we have just described for the time-evolution and evaluation of matrix elements corresponds to that used in the Schwinger-Keldysh formalism \cite{sk}.  This section briefly reviews enough of this formalism to make clear our subsequent calculations; a fuller account appears in the appendix of \cite{greens}.

To begin, consider the time evolution to be that given by the interaction picture.  There, the evolution of operators is determined by the free part of the Hamiltonian; for the field $\varphi(x)$, this time-dependence is already precisely that contained in the modes since they are the solutions of the free Klein-Gordon equation.  The evolution of the state is set by the interacting part of the Hamiltonian, $H_I(\eta)$.  If we define a time-evolution operator $U_I(\eta,\eta_0)$ so that 
\begin{equation}
|0_{\rm eff}(\eta)\rangle = U_I(\eta,\eta_0)\, |0_{\rm eff}(\eta_0)\rangle
\equiv U_I(\eta,\eta_0)\, |0_{\rm eff}\rangle , 
\label{timeevolve}
\end{equation}
then this operator is given by Dyson's formula, 
\begin{equation}
U_I(\eta,\eta_0) = T e^{-i \int_{\eta_0}^\eta d\eta'\, H_I(\eta')} . 
\label{dyson}
\end{equation}
Note that whenever we write the state without its time argument, it is implicitly evaluated at the initial time, 
\begin{equation}
|0_{\rm eff}\rangle = |0_{\rm eff}(\eta_0)\rangle . 
\label{noinddef}
\end{equation}

Besides starting at a particular time, rather than at some asymptotically distant past, another distinguishing feature of the Schwinger-Keldysh approach is that both the state and its dual are evolved, 
\begin{equation}
\langle 0_{\rm eff}(\eta) | {\cal O}(\eta) | 0_{\rm eff}(\eta) \rangle 
= \langle 0_{\rm eff} | U_I^\dagger(\eta,\eta_0) {\cal O}(\eta) U_I(\eta,\eta_0)| 0_{\rm eff} \rangle 
\label{matrixUdOU}
\end{equation}
The appearance of two separately time-ordered operators, $U_I$ and $U_I^\dagger$, is a little cumbersome, so the next sequence of steps introduces a more compact notation which keeps this time-ordering intact.  

We first insert the identity operator in the form, $U_I^\dagger(0,\eta)U_I(0,\eta)$, so that Eq.~(\ref{matrixUdOU}) becomes 
\begin{equation}
\langle 0_{\rm eff} | U_I^\dagger(0,\eta_0) U_I(0,\eta) {\cal O}(\eta) U_I(\eta,\eta_0)| 0_{\rm eff} \rangle 
\label{matrixUdUdUOU}
\end{equation}
where we have simplified, $U_I^\dagger(\eta,\eta_0) U_I^\dagger(0,\eta) = U_I^\dagger(0,\eta_0)$.  Reading the operator in Eq.~(\ref{matrixUdUdUOU}) from right to left, we follow the time evolution from $\eta_0$ to $\eta$, where the operator is inserted, then continue to the asymptotic future\footnote{In conformal coordinates, $\eta$ is usually taken to run from $-\infty$ (the asymptotic past) to $0$ (the asymptotic future).} before then proceeding backwards---because of the Hermitian conjugation in $U_I^\dagger(0,\eta_0)$---from $0$ back to the initial time $\eta_0$.  Since the first three operators along this path are already correctly ordered, let us label their arguments with a `$+$' while the arguments of the last operator, $U_I^\dagger(0,\eta_0)$, are labeled with a `$-$.'  This notation allows us to group together all of the exponentials within a single time-ordering symbol, 
\begin{equation}
\langle 0_{\rm eff} | 
T \bigl( 
{\cal O}(\eta^+) 
e^{-i\int_{\eta_0}^0 d\tilde\eta^+\, H_I(\tilde\eta^+)} 
e^{-i\int^{\eta_0}_0 d\tilde\eta^-\, H_I(\tilde\eta^-)} 
\bigr) | 0_{\rm eff} \rangle , 
\label{matrixTexppm}
\end{equation}
provided we recall that times on the `$-$' part of the path occur {\it after\/} and {\it in the opposite order\/} as those on the `$+$' part of the path.

Finally, it is most convenient to exchange the doubling of the time coordinate for a doubling of the field content; $\varphi^+$ and $h_{\mu\nu}^+$ fields are therefore always implicitly on the `$+$' part of the path whereas $\varphi^-$ and $h_{\mu\nu}^-$ are always on the `$-$' part.  With this change, the index on the conformal time is superfluous and hereafter will not be written.  The time-evolution of a general matrix element is then very compactly written as 
\begin{equation}
\langle 0_{\rm eff} | 
T \bigl( 
{\cal O}^+(\eta) 
e^{-i\int_{\eta_0}^0 d\eta'\, [ H_I^+(\eta') - H_I^-(\eta') ]} 
\bigr) | 0_{\rm eff} \rangle , 
\label{matrixTHpm}
\end{equation}
where $H_I^\pm$ and ${\cal O}^+$ refer to operators composed entirely of $+$ or $-$ fields as indicated.

When we expand the time-ordered operator in a perturbative evaluation of the matrix element, we encounter four distinct propagators depending upon which pairs of fields are being Wick contracted.  As an example, for the scalar field we have
\begin{equation}
-i G^{\pm\pm}(x,x') = \langle 0_{\rm eff} | T \bigl( \varphi^\pm(x) \varphi^\pm(x') \bigr) | 0_{\rm eff} \rangle 
\label{Gpmpm}
\end{equation}
or 
\begin{equation}
G^{\pm\pm}(x,x') = \int {d^3\vec k\over (2\pi)^3}\, e^{i\vec k\cdot(\vec x-\vec x')}\, G_k^{\pm\pm}(\eta,\eta') . 
\label{Gkpmpm}
\end{equation}
Since the initial-state structure is defined at the earliest possible time in this framework, the effective state part of all these propagators is the same and the only differences lie in the the standard, point-source parts of the propagators.  The $\Theta$-functions that appear in these parts are inherited directly from the time-ordering implied by the ordering of the $\varphi^+$ and $\varphi^-$ fields on the time contour, 
\begin{eqnarray}
-iG_k^{++}(\eta,\eta') &\!\!\!=\!\!\!& 
\Theta(\eta-\eta')\, U_k(\eta) U_k^*(\eta') 
\nonumber \\
&& 
+ \Theta(\eta'-\eta)\, U_k^*(\eta) U_k(\eta') 
\nonumber \\
&& 
+ f_k^* U_k(\eta) U_k(\eta') 
\nonumber \\
-iG_k^{+-}(\eta,\eta') &\!\!\!=\!\!\!& 
U_k^*(\eta) U_k(\eta') 
+ f_k^* U_k(\eta) U_k(\eta') 
\nonumber \\
-iG_k^{-+}(\eta,\eta') &\!\!\!=\!\!\!& 
U_k(\eta) U_k^*(\eta') 
+ f_k^* U_k(\eta) U_k(\eta') 
\nonumber \\
-iG_k^{--}(\eta,\eta') &\!\!\!=\!\!\!& 
\Theta(\eta'-\eta)\, U_k(\eta) U_k^*(\eta') 
\nonumber \\
&& 
+ \Theta(\eta-\eta')\, U_k^*(\eta) U_k(\eta') 
\nonumber \\
&& 
+ f_k^* U_k(\eta) U_k(\eta') . 
\label{Gpmpmexpl}
\end{eqnarray}
Thus, for example, $G_k^{--}(\eta,\eta')$ has exactly the opposite time-ordering as $G_k^{++}(\eta,\eta')$ since the $\varphi^-$ fields came from the $U_I^\dagger(0,\eta_0)$ operator whose time-ordering is reversed due to the Hermitian conjugation.

An analogous structure also applies to the graviton propagator, 
\begin{equation}
-i \Pi^{\pm\pm}_{\mu\nu,\lambda\rho}(x,x') \equiv \langle 0_{\rm eff} | 
T \bigl( h_{\mu\nu}^\pm(x) h_{\lambda\rho}^\pm(x') \bigr) | 0_{\rm eff} \rangle
\label{hpropdef}
\end{equation}
with 
\begin{equation}
\Pi^{\pm\pm}_{\mu\nu,\lambda\rho}(x,x') 
= \int {d^3\vec k\over (2\pi)^3}\, e^{i\vec k\cdot(\vec x-\vec x')} 
\Pi^{\pm\pm}_{\mu\nu,\lambda\rho}(\eta,\eta';\vec k) . 
\label{hpropk}
\end{equation}
These propagators are determined by three basics tensors whose structure parallels that of the scalar field propagators, 
\begin{eqnarray}
\Pi^{++}_{\mu\nu,\lambda\rho}(\eta,\eta';\vec k) &\!\!\!=\!\!\!& 
\Theta(\eta-\eta')\, \Pi^>_{\mu\nu,\lambda\rho}(\eta,\eta';\vec k)
\nonumber \\
&& 
+ \Theta(\eta'-\eta)\, \Pi^<_{\mu\nu,\lambda\rho}(\eta,\eta';\vec k) 
\nonumber \\
&&
+ \tilde\Pi_{\mu\nu,\lambda\rho}(\eta,\eta';\vec k) 
\nonumber \\
\Pi^{+-}_{\mu\nu,\lambda\rho}(\eta,\eta';\vec k) &\!\!\!=\!\!\!& 
\Pi^<_{\mu\nu,\lambda\rho}(\eta,\eta';\vec k) 
+ \tilde\Pi_{\mu\nu,\lambda\rho}(\eta,\eta';\vec k) 
\nonumber \\
\Pi^{-+}_{\mu\nu,\lambda\rho}(\eta,\eta';\vec k) &\!\!\!=\!\!\!& 
\Pi^<_{\mu\nu,\lambda\rho}(\eta,\eta';\vec k) 
+ \tilde\Pi_{\mu\nu,\lambda\rho}(\eta,\eta';\vec k) 
\nonumber \\
\Pi^{--}_{\mu\nu,\lambda\rho}(\eta,\eta';\vec k) &\!\!\!=\!\!\!& 
\Theta(\eta'-\eta)\, \Pi^>_{\mu\nu,\lambda\rho}(\eta,\eta';\vec k) 
\nonumber \\
&& 
+ \Theta(\eta-\eta')\, \Pi^<_{\mu\nu,\lambda\rho}(\eta,\eta';\vec k) 
\nonumber \\
&&
+ \tilde\Pi_{\mu\nu,\lambda\rho}(\eta,\eta';\vec k) . 
\label{Gpmpmforh}
\end{eqnarray}
Here we have written the state-dependent part of the propagator as $\tilde\Pi_{\mu\nu,\lambda\rho}(\eta,\eta';\vec k)$.  This tensor can be written, as in the case of the scalar field, as products of Bunch-Davies modes multiplied by an initial state structure function.  In principle, we could have separate structure functions for the scalar, vector and tensor modes, depending upon the symmetries of the initial state.  However, since in the processes we shall examine the graviton propagator only appears as a common external leg in each of a set of diagrams, it will not be necessary to have an explicit expressions for either $\Pi^{>,<}_{\mu\nu,\lambda\rho}(\eta,\eta';\vec k)$ or $\tilde\Pi_{\mu\nu,\lambda\rho}(\eta,\eta';\vec k)$.

\section{The graviton tadpole}
\label{graviton}

Having described generally how the information within the effective initial state propagates forward, we are ready to evaluate the loop correction to the graviton tadpole.  This case is a specific instance of the formalism that we have just described, with ${\cal O}^+ = h_{\mu\nu}^+(\eta,\vec x)$ and with an interaction Hamiltonian derived from the action presented earlier in Eq.~(\ref{interaction}), 
\begin{equation}
H_I(\eta) = {1\over 2} a^2(\eta) \int d^3\vec x\, 
h^{\mu\nu} \bigl\{
- 2 \tilde G_{\mu\nu} + T^{\rm cl}_{\mu\nu} + \hat T_{\mu\nu} \bigr\} . 
\label{HIdef}
\end{equation}
The first two terms to which the graviton couples are purely classical quantities, for the background geometry and the zero mode of the scalar field, while the last piece is quadratic in the quantum fluctuations of the scalar field. 

\begin{figure}[!tbp]
\begin{center}
\includegraphics{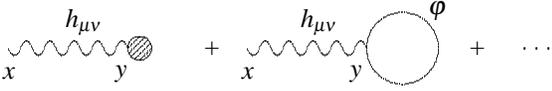}
\caption{The leading contributions to the graviton tadpole condition.  The shaded blob represents the coupling of the graviton to the classical background tensor, $-2\tilde G_{\mu\nu} + T_{\mu\nu}^{\rm cl}$. 
\label{htadpole}}
\end{center}
\end{figure}

Evaluating the graviton tadpole to first order in this interaction yields
\begin{eqnarray}
&&\!\!\!\!\!\!\!
\langle 0_{\rm eff}(\eta) | h^+_{\mu\nu}(\eta,\vec x) | 0_{\rm eff}(\eta) \rangle 
\nonumber \\
&\!\!\!=\!\!\!& 
\langle 0_{\rm eff} | T \bigl( h^+_{\mu\nu}(\eta,\vec x) e^{-i\int_{\eta_0}^0 d\eta' [H_I^+(\eta') - H_I^-(\eta')]} \bigr) | 0_{\rm eff} \rangle 
\nonumber \\
&\!\!\!=\!\!\!& 
{1\over 2} \int_{\eta_0}^\eta d\eta' a^2(\eta') 
\Bigl\{
\bigl[ \Pi^{>\ \, \lambda\rho}_{\mu\nu,}(\eta,\eta';\vec 0) 
- \Pi^{<\ \, \lambda\rho}_{\mu\nu,}(\eta,\eta';\vec 0) \bigr]
\nonumber \\
&&\qquad 
\times\bigl[
2 \tilde G_{\lambda\rho}(\eta') - T^{\rm cl}_{\lambda\rho}(\eta') 
- T_{\lambda\rho}(\eta')  \bigr]
+ \cdots \Bigr\} . 
\label{tadpoleeval}
\end{eqnarray}
Grouping together the first two classical contributions in the integrand, these terms in the tadpole can be represented diagrammatically as in Fig.~\ref{htadpole}.

The contribution from the the scalar fluctuations $T_{\mu\nu}$ to the energy-momentum contains divergences and this section will show how they are absorbed by rescaling the parameters in the contribution from the classical background, $\tilde G_{\mu\nu}$.  The divergences in $T_{\mu\nu}$ fall naturally into two classes.  The first occur at arbitrary times during the evolution, so we shall call such infinities bulk divergences, and they are independent of the structure of the initial state.  The second class is closely associated with how we have chosen the effective initial state.  Since these divergences only occur at the initial time \cite{emtensor}, $\eta=\eta_0$, we shall characterize this set as boundary divergences.  The treatment of the bulk divergences has been well established \cite{emtensor,books}, so we include only a cursory discussion of them here and shall examine the boundary divergences in much more detail.

In this setting, the standard gravitational equations of motion are imposed by the vanishing of the tadpole in Eq.~(\ref{tadpoleeval}), which is satisfied when the integrand vanishes, 
\begin{equation}
 T_{\mu\nu} + T^{\rm cl}_{\mu\nu} = 2 \tilde G_{\mu\nu} . 
\label{bulkeom}
\end{equation}
In our isotropically expanding background, this equation implies two others for its temporal, 
\begin{eqnarray}
\rho + \rho_0 
&\!\!\!=\!\!\!& 
2 \Lambda + 6 M^2_{\rm pl} {1\over a^2} \biggl( {a'\over a} \biggr)^2 
\label{bulkeom00} \\
&& 
+ {36\alpha\over a^4} \biggl[ 
2 {a^{\prime\prime\prime}\over a} {a'\over a} 
- \biggl( {a^{\prime\prime}\over a} \biggr)^2 
- 4 {a^{\prime\prime}\over a} \biggl( {a'\over a} \biggr)^2 
\biggr] , 
\nonumber 
\end{eqnarray}
and its spatial components, 
\begin{eqnarray}
p + p_0 
&\!\!\!=\!\!\!& 
- 2 \Lambda 
- 2 M^2_{\rm pl} {1\over a^2} \biggl[ 
2 {a^{\prime\prime}\over a} - \biggl( {a'\over a} \biggr)^2 
\biggr] 
\label{bulkeomij} \\
&& 
- {24\alpha\over a^4} \biggl[ 
{a^{\prime\prime\prime\prime}\over a} 
- 5 {a^{\prime\prime\prime}\over a} {a'\over a} 
- {5\over 2} \biggl( {a^{\prime\prime}\over a} \biggr)^2 
+ 8 {a^{\prime\prime}\over a} \biggl( {a'\over a} \biggr)^2 
\biggr] . 
\nonumber 
\end{eqnarray}
Here, $\rho$ and $p$ are as given in Eqs.~(\ref{rhoval})--(\ref{pval}) and Eq.~(\ref{Gtildedef}) provided the gravitational part.  The contributions from the zero mode of the scalar field has been abbreviated by  
\begin{eqnarray}
a^2\, \rho_0 &\!\!\!=\!\!\!& 
{\textstyle{1\over 2}} (\phi')^2 + {\textstyle{1\over 2}} a^2m^2 \phi^2 
\nonumber \\
a^2\, p_0 &\!\!\!=\!\!\!& 
{\textstyle{1\over 2}} (\phi')^2 - {\textstyle{1\over 2}} a^2m^2 \phi^2 . 
\label{rhopdef0}
\end{eqnarray}
We now look at each type of divergence separately.

\subsection{Bulk renormalization}

The divergences that require the renormalization of the parameters of the bulk gravitational action, $\Lambda$, $M_{\rm pl}$ and $\alpha$, are produced by the part of the density and pressure that is independent of the initial-state structure, 
\begin{eqnarray}
\rho_{\rm bulk} &\!\!\!=\!\!\!& 
{1\over 2} {1\over a^2} \int {d^3\vec k\over (2\pi)^3}\, 
\bigl[ U_k' U_k^{\prime *} + (k^2 + a^2m^2) U_k U_k^* \bigr] 
\nonumber \\
p_{\rm bulk} &\!\!\!=\!\!\!& - \rho_{\rm bulk}  
+ {1\over a^2} \int {d^3\vec k\over (2\pi)^3}\, 
\Bigl[ U_k' U_k^{\prime *} + {\textstyle{1\over 3}} k^2 U_k U_k^* \Bigr] . 
\qquad
\label{bulkrhop}
\end{eqnarray}
Each of these expressions contains quartic, quadratic and logarithmic divergences which occur at any conformal time.  Since the same renormalization absorbs the divergences for both the pressure and the density, it is only necessary to show the case of the density explicitly.  

To isolate these divergences from the finite parts of the energy-momentum, we apply our adiabatic expansion given in Eq.~(\ref{Omegaasns}) and then further expand the generalized frequency $\Omega_k(\eta)$ in inverse powers of $\omega_k=\sqrt{k^2+a^2m^2}$, as in Eq.~(\ref{Omegato5}), 
\begin{eqnarray}
&&\!\!\!\!\!\!
\rho_{\rm bulk}
\nonumber \\
&\!\!\!=\!\!\!&
{1\over 2} {1\over a^4} \int {d^3\vec k\over (2\pi)^3}\, \omega_k 
+ {1\over 4} {1\over a^4} \biggl( {a'\over a} \biggr)^2 
\int {d^3\vec k\over (2\pi)^3}\, 
\biggl[ {1\over\omega_k} + {a^2 m^2\over\omega_k^3} \biggr] 
\nonumber \\
&& 
-\ {1\over 16} {1\over a^4} \biggl[
2 {a^{\prime\prime\prime}\over a} {a'\over a}
- \biggl( {a^{\prime\prime}\over a} \biggr)^2
- 4 {a^{\prime\prime}\over a} \biggl( {a'\over a} \biggr)^2
\biggr] 
\int {d^3\vec k\over (2\pi)^3}\, {1\over\omega_k^3} 
\nonumber \\
&&
+ {\rm finite} . 
\label{rhodiv}   
\end{eqnarray}
Although the integrals are only over the three spatial dimensions, their structures are otherwise very similar to the momentum integrals ordinarily encountered in $S$-matrix calculations once we dimensionally regularize them and extend the number of dimensions to $3-2\epsilon$, 
\begin{eqnarray}
I^{(n)}(\epsilon) &\!\!\!=\!\!\!& 
\int {d^{3-2\epsilon}\vec k\over (2\pi)^{3-2\epsilon}}\, 
{(a\mu)^{2\epsilon}\over\omega_k^n} 
\nonumber \\
&\!\!\!=\!\!\!& 
{\sqrt{\pi}\over 8\pi^2} {\Gamma\bigl( \epsilon - {3-n\over 2} \bigr)\over \Gamma\bigl( {n\over 2} \bigr)} 
\biggl[ {4\pi\mu^2\over m^2} \biggr]^\epsilon (am)^{3-n} . 
\label{dr}
\end{eqnarray}
Applying this general integral to each of the specific divergences in Eqs.~(\ref{rhodiv}) and restoring the $\epsilon\to 0$ limit we find the following poles, 
\begin{eqnarray}
\rho_{\rm bulk}&\!\!\!=\!\!\!&
- {m^4\over 64\pi^2} {1\over\epsilon} 
+ {m^2\over 192\pi^2} {6\over a^2} \biggl( {a'\over a} \biggr)^2 
{1\over\epsilon} 
\nonumber \\
&& 
- {1\over 2304\pi^2} {36\over a^4} \biggl[
2 {a^{\prime\prime\prime}\over a} {a'\over a}
- \biggl( {a^{\prime\prime}\over a} \biggr)^2
- 4 {a^{\prime\prime}\over a} \biggl( {a'\over a} \biggr)^2 \biggr] 
{1\over\epsilon}
\nonumber \\
&&
+ {\rm finite} . 
\label{rhopole}   
\end{eqnarray}

The prefactors that multiply each pole have precisely the same structures, in their dependence on the scale factor and its derivatives, as the terms appearing in $\tilde G_{00}$ defined in Eq.~(\ref{Gtildedef}).  This equivalence allows us to cancel the divergences, which arose from the energy-momentum contained in the short-distance properties of the scalar field, through an appropriate redefinition of the gravitational parameters $\Lambda$, $M_{\rm pl}$ and $\alpha$.  Minimally, we could define, 
\begin{eqnarray}
\Lambda_R &\!\!\!=\!\!\!& 
\Lambda + {m^4\over 64\pi^2} {1\over\epsilon} 
\nonumber \\
M_{{\rm pl},R}^2 &\!\!\!=\!\!\!& 
M_{\rm pl}^2 - {m^2\over 192\pi^2} {1\over\epsilon}
\nonumber \\
\alpha_R &\!\!\!=\!\!\!& 
\alpha + {1\over 2304\pi^2} {1\over\epsilon} , 
\label{bulkMS}    
\end{eqnarray}
although we could also apply an $\overline{\rm MS}$ scheme to remove some of the artifacts of dimensional regularization which were not shown explicitly here.  The renormalized parameters also acquire a dependence on the renormalization scale $\mu$, as is explained more fully in \cite{emtensor}.

\subsection{Boundary renormalization}

Although we have separated the density and the pressure into ``bulk'' and ``surface'' components, it is important to remember that both properly contribute to the bulk equations of motion implied by the vanishing of the graviton tadpole, 
\begin{eqnarray}
\rho_{\rm bulk} + \rho_{\rm surf} + \rho_0 &\!\!\!=\!\!\!& 
2\Lambda_R + {6M_{{\rm pl},R}^2\over a^2} \biggl( {a'\over a} \biggr) 
+ \cdots 
\nonumber \\
p_{\rm bulk} + p_{\rm surf} + p_0 &\!\!\!=\!\!\!& 
- 2\Lambda_R - {2M_{{\rm pl},R}^2\over a^2} \biggl[ 
2 {a^{\prime\prime}\over a} - \biggl( {a'\over a} \biggr) \biggr]
\nonumber \\
&&
+ \cdots . 
\label{bulkeom2}
\end{eqnarray} 
What is unique about the surface, or state-dependent, components of the energy-momentum tensor is that the divergences they produce are confined to the initial time, $\eta'=\eta_0$, unlike those of the bulk, or state-independent, component which occur for any value of $\eta'$.  Thus these new divergences cannot be absorbed through a redefinition of the bulk properties of the theory, but rather are removed by adding the appropriate gravitational counterterms confined to the initial boundary.  

To discover when these divergences occur and how they are cancelled, in this section we concentrate only on the contribution to the graviton tadpole provided by $\rho_{\rm surf}$ and $p_{\rm surf}$, 
\begin{eqnarray}
0 &\!\!\!=\!\!\!&
\langle 0_{\rm eff}(\eta) | h_{\mu\nu}^+(\eta,\vec x) | 0_{\rm eff} \rangle 
\nonumber \\
&\!\!\!=\!\!\!&
- {1\over 2} \int_{\eta_0}^\eta d\eta'\, \bigl\{ 
{\cal G}_{\mu\nu,}^{\ \ 00}(\eta,\eta')\, \rho_{\rm surf}(\eta') 
\nonumber \\
&&\qquad\qquad\quad
+ {\cal G}_{\mu\nu,}^{\ \ ij}(\eta,\eta')\delta_{ij}\, p_{\rm surf}(\eta') 
+ \cdots \bigr\} . \qquad
\label{surftadpole}
\end{eqnarray}
Here we have abbreviated the external graviton leg by the function, 
\begin{equation}
{\cal G}_{\mu\nu, \lambda\rho}(\eta,\eta')
\equiv a^4(\eta')\, \bigl[ 
\Pi^>_{\mu\nu, \lambda\rho}(\eta,\eta';\vec 0)
- \Pi^<_{\mu\nu, \lambda\rho}(\eta,\eta';\vec 0) \bigr] , 
\label{externalleg}
\end{equation}
since it occurs frequently in the analysis of the boundary divergences.

The state-dependent part of the energy-momentum tensor can diverge at the initial surface whenever the difference between the nominal and the true vacuum states grows sufficiently fast at short distances.  The general method for isolating these divergences, for a particular choice of the initial state's structure, will be first to determine which terms within $\rho_{\rm surf}$ and $p_{\rm surf}$ produce the divergences and then to integrate these terms by parts sufficiently many times until the integrand in Eq.~(\ref{surftadpole}) is finite.  In the process we generate terms explicitly evaluated on the initial boundary that diverge when we sum over all momentum scales.  Such divergences can be regulated by continuing the number of dimensions in the spatial momentum integrals of these boundary terms.  Once we have extracted the poles explicitly, we renormalize the theory by adding the appropriate counterterms on the boundary.

This method for isolating and renormalizing the boundary divergences is best illustrated through a sequence of simple examples, which we present in the next section, but we first describe some of the general properties of this method.  Because the generalized frequency appears in the oscillating factor, $e^{-2i\int_{\eta_0}^\eta d\eta'\, \Omega_k(\eta')}$, that is present in $\rho_{\rm surf}$ and $p_{\rm surf}$, 
\begin{eqnarray}
\rho_{\rm surf} &\!\!\!=\!\!\!& 
{1\over 4} {1\over a^4} \int {d^3\vec k\over (2\pi)^3}\, 
f_k^* e^{-2i\int_{\eta_0}^\eta d\eta'\, \Omega_k(\eta')}
\nonumber \\
&&\times 
\biggl\{ 2i {a'\over a} 
+ {1\over\Omega_k} \biggl[ {a^{\prime\prime}\over a} + \biggl( {a'\over a} \biggr)^2 \biggr] 
+ i {\Omega'_k\over\Omega_k}
\nonumber \\
&&\quad 
+ {a'\over a} {\Omega'_k\over\Omega_k^2} 
+ {1\over 2} {\Omega_k^{\prime\prime}\over\Omega_k^2} 
- {1\over 2} {\Omega_k^{\prime\, 2}\over\Omega_k^3} 
\biggr\}
\nonumber \\
p_{\rm surf} &\!\!\!=\!\!\!& 
{1\over 4} {1\over a^4} \int {d^3\vec k\over (2\pi)^3}\, 
f_k^* e^{-2i\int_{\eta_0}^\eta d\eta'\, \Omega_k(\eta')}
\nonumber \\
&&\times 
\biggl\{ - {4\over 3} \Omega_k + 2i {a'\over a} 
- {1\over 3} {1\over\Omega_k} \biggl[ {a^{\prime\prime}\over a} - 3 \biggl( {a'\over a} \biggr)^2 + 2 a^2m^2 \biggr] 
\nonumber \\
&&\quad 
+ i {\Omega'_k\over\Omega_k}
+ {a'\over a} {\Omega'_k\over\Omega_k^2} 
- {1\over 6} {\Omega_k^{\prime\prime}\over\Omega_k^2} 
+ {1\over 2} {\Omega_k^{\prime\, 2}\over\Omega_k^3} 
\biggr\} , 
\label{rhopOmega}
\end{eqnarray}
we shall express the terms that compose the density and pressure in terms of $\Omega_k$.  Actually, since the structure of the effective state is defined at the initial boundary, a general term in the energy-momentum tensor contains factors of both $\Omega_k(\eta)$ and $\Omega_k(\eta_0)$, using Eq.~(\ref{Omegato5}) for example to convert as necessary the factors of $\omega_k(\eta_0)$ in $f_k$ as we defined it in Eq.~(\ref{fkdef}).  Based on such observations, it is natural to define the following kernel function, 
\begin{equation}
K^{(p)}(\eta) = \int {d^3\vec k\over (2\pi)^3}\, 
{e^{-2i\int_{\eta_0}^\eta d\eta'\, \Omega_k(\eta')}\over \Omega_k^{3-p}(\eta) \Omega_k^p(\eta_0) } , 
\label{kerneldef}
\end{equation}
which diverges logarithmically as $\eta$ approaches $\eta_0$ \cite{greens,emtensor}, or logarithmically in a cutoff, applied to large spatial momenta, when evaluated on the initial surface.

We can now explain in more detail our approach for isolating the boundary divergences.  We first write all the divergent parts of $\rho_{\rm surf}$ and $p_{\rm surf}$ in terms of the kernel $K^{(p)}(\eta)$ or its derivatives.  Since the kernel only diverges logarithmically, the overall time integral still remaining in the graviton tadpole means that it gives a finite contribution to the matrix element.  The goal then is to integrate these derivatives of the kernel until we have moved them to the external leg, generating a series of surface terms explicitly evaluated at the initial boundary.  These surface terms no longer contain the oscillatory factor in Eq.~(\ref{kerneldef}), so the remaining divergent spatial momentum integrals can be readily dimensionally regulated using Eq.~(\ref{dr}).

Once we have extracted and regularized the boundary divergences, we renormalize the theory by adding counterterms which are similarly confined to the initial surface.  For the small fluctuations we have been considering, we require only the part of the counterterm Hamiltonian that is linear in the graviton, $h_{\mu\nu}$, 
\begin{equation}
H_I^{\rm c.t.}(\eta) = - \int d^3\vec x\, \bigl\{ 
a^3(\eta) h^{\mu\nu} \delta T_{\mu\nu}(\eta)\, \delta(\eta-\eta_0) 
\bigr\} . 
\label{bndct3}
\end{equation}
where the $\delta$-function (and its derivatives) confine its effects to the boundary.  Here, $\delta T_{\mu\nu}$ is a purely classical geometric tensor, that is, it depends on the scale factor and its derivatives but not the scalar field and it can also contain derivatives that act on the $\delta$-function.  The specific form of $\delta T_{\mu\nu}$ depends on the particular initial state chosen and we illustrate its form for a few examples.

\section{Three simple examples}
\label{examples}

The technique for boundary renormalization that we have outlined in the previous section can be straightforwardly applied to any of the terms in our power series expansion of $f_k$, appearing in Eq.~(\ref{fkdef}).  Since there is little to distinguish one case from another beyond its analytic complexity, we shall show the three simplest examples of boundary renormalization here.  An additional advantage of studying these cases is that their geometric interpretation is most transparent---for instance, the first divergence requires a boundary counterterm which is essentially a surface tension, which is analogous to the cosmological constant that appeared in the bulk renormalization.

\subsection{The surface tension}

Among the terms contained in the first sum in Eq.~(\ref{fkdef}) that describes $f_k$---those that scale as an inverse power of $\omega_k$---very few lead to divergences in the energy-momentum tensor since the departures from the Bunch-Davies state diminishes at shorter distances for these moments.  Only the first three terms in this series lead to boundary divergences and the counterterms associated with them correspond to dimension zero, one and two relevant operators on the boundary.  

The unique dimension zero counterterm is the surface tension.  We begin with this case by considering an initial state whose structure is given by 
\begin{equation}
f_k = c_3 {a^3(\eta_0) m^3\over\omega_k^3(\eta_0)} . 
\label{fk3eg}
\end{equation}
The potentially divergent parts of the state-dependent components of the density and pressure are 
\begin{eqnarray}
\rho_{\rm surf} &\!\!\!=\!\!\!& 
{c_3^* m^3\over 4} {a^3(\eta_0)\over a^4(\eta')} \int {d^3\vec k\over (2\pi)^3}\, 
{e^{-2i\int_{\eta_0}^{\eta'}d\eta^{\prime\prime}\, \Omega_k(\eta^{\prime\prime})}\over \omega_k^3(\eta_0)} 
\nonumber \\
&&\qquad\quad\times 
\biggl[ 2i {a'(\eta')\over a(\eta')} 
+ {\cal O} \biggl( {1\over\omega_k(\eta')} \biggr) \biggr]
\nonumber \\
p_{\rm surf} &\!\!\!=\!\!\!& 
{c_3^* m^3\over 4} {a^3(\eta_0)\over a^4(\eta')} \int {d^3\vec k\over (2\pi)^3}\, 
{e^{-2i\int_{\eta_0}^{\eta'}d\eta^{\prime\prime}\, \Omega_k(\eta^{\prime\prime})}\over \omega_k^3(\eta_0)} 
\nonumber \\
&&\qquad\quad\times 
\biggl[ - {4\over 3} \omega_k(\eta') 
+ 2i {a'(\eta')\over a(\eta')} 
+ {\cal O} \biggl( {1\over\omega_k(\eta')} \biggr) \biggr] . 
\nonumber \\
\label{eg3rhop}
\end{eqnarray}
Both the leading divergence in the density and the sub-leading divergence in the pressure grow near the boundary only as $\ln(\eta'-\eta_0)$ \cite{emtensor}.  Such a divergence is integrable so, given that we still have a conformal time integration in the graviton tadpole, these terms yield a finite contribution to Eq.~(\ref{surftadpole}).  Thus, only the pressure diverges.

To treat the remaining term in the pressure, let us consider the first derivative of the kernel function that we introduced in Eq.~(\ref{kerneldef}), for the case $p=3$, 
\begin{equation}
K^{(3)\prime}(\eta) = -2i \int {d^3\vec k\over (2\pi)^3}\, 
{\omega_k(\eta) e^{-2i\int_{\eta_0}^\eta d\eta'\, \Omega_k(\eta')}\over 
\omega_k^3(\eta_0)} + {\rm finite} ; 
\label{kernel4eg}
\end{equation}
this expression has exactly the same structure as the divergent term in the pressure, up to finite corrections.  Therefore, we write the divergent contribution to the graviton tadpole as  
\begin{eqnarray}
&&\!\!\!\!\!\!\!
\langle 0_{\rm eff}(\eta) | h_{\mu\nu}^+(\eta,\vec x)|0_{\rm eff}(\eta) \rangle
\label{eg3div} \\ 
&\!\!\!=\!\!\!& {ic_3^* m^3\over 12} a^3(\eta_0) \int_{\eta_0}^\eta d\eta'\, \biggl\{ 
{\cal G}_{\mu\nu,}^{\ \ ij}(\eta,\eta')\delta_{ij}\, 
{K^{(3)\prime}(\eta') \over a^4(\eta')} 
+ \cdots \biggr\} . 
\nonumber
\end{eqnarray}
Integrating the parts, we remove the derivative from the kernel to obtain an explicitly finite integral, since its integrand now only diverges logarithmically, as well as a divergent contribution which is now explicitly evaluated on the initial boundary, 
\begin{eqnarray}
&&\!\!\!\!\!\!\!
\langle 0_{\rm eff}(\eta) | h_{\mu\nu}^+(\eta,\vec x)|0_{\rm eff}(\eta) \rangle
\nonumber \\
&\!\!\!=\!\!\!& 
- {ic_3^* m^3\over 12} 
{\cal G}_{\mu\nu,}^{\ \ ij}(\eta,\eta_0)\delta_{ij}\, 
{1\over a(\eta_0)} K^{(3)}(\eta_0)  
\nonumber \\
&& 
- {ic_3^* m^3\over 12} a^3(\eta_0) \int_{\eta_0}^\eta d\eta'\, \biggl\{ 
K^{(3)\prime}(\eta') {\partial\over\partial\eta'} \biggl[  
{{\cal G}_{\mu\nu,}^{\ \ ij}(\eta,\eta')\delta_{ij}\over a^4(\eta')} 
\biggr] \biggr\} 
\nonumber \\
&& + \cdots . 
\label{eg3ipb}
\end{eqnarray}
When the kernel is evaluated on the boundary, the oscillating factor disappears and the terms in the denominator coalesce to become 
\begin{equation}
K^{(p)}(\eta_0) = \int {d^3\vec k\over (2\pi)^3}\, 
{1\over\Omega_k^3(\eta_0)} 
= \int {d^3\vec k\over (2\pi)^3}\, 
{1\over\omega_k^3(\eta_0)} + \cdots ,  
\label{kernelbnd}
\end{equation}
We find the pole by dimensionally regulating the remaining momentum integral, 
\begin{equation}
K^{(p)}(\eta_0) 
= {1\over 4\pi^2} \biggl[ {1\over\epsilon} - \gamma + \ln {4\pi\mu^2\over m^2} \biggr] 
+ \cdots . 
\label{kernelpole}
\end{equation}
When we substitute this result back into the graviton tadpole, we obtain the following boundary divergence, 
\begin{eqnarray}
&&\!\!\!\!\!\!\!
\langle 0_{\rm eff}(\eta) | h_{\mu\nu}^+(\eta,\vec x)|0_{\rm eff}(\eta) \rangle
\nonumber \\
&\!\!\!=\!\!\!& - {ic_3^* m^3\over 48\pi^2} 
{{\cal G}_{\mu\nu,}^{\ \ ij}(\eta,\eta_0)\delta_{ij}\over a(\eta_0)} 
\biggl[ {1\over\epsilon} - \gamma + \ln {4\pi\mu^2\over m^2} \biggr] 
+ \cdots . \qquad
\label{eg3divple}
\end{eqnarray}

The meaning of this term can be understood through its structure.  The external propagator leg, represented by the function ${\cal G}(\eta,\eta_0)$, connects a general time $\eta$ with the initial time $\eta_0$.  When we sum over arbitrary intermediate space-time points, such as the internal vertex of the loop diagram in Fig.~\ref{htadpole}, we are also summing over all the short-distance structures of the state whenever that intermediate point lies on the initial boundary.   To cancel the divergence associated with this sum, we must add an interaction counterterm that is also confined to this initial surface, which thereby guarantees that the external leg vanishes except when the intermediate vertex lies on this surface.  This construction ensures that the counterterm contribution to the tadpole has correct conformal time dependence to cancel pole in Eq.~(\ref{eg3divple}).  
\begin{figure}[!tbp]
\includegraphics{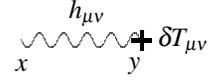}
\caption{The contribution from a surface counterterm insertion to the graviton tadpole.  
\label{cttadpole}}
\end{figure}

The generic form for the linear part of the surface counterterm in the fluctuation is given in Eq.~(\ref{bndct3}) and its contribution is shown in Fig.~\ref{cttadpole}.  Adding its effect to the boundary divergence yields 
\begin{eqnarray}
&&\!\!\!\!\!\!\!
\langle 0_{\rm eff}(\eta) | h_{\mu\nu}^+(\eta,\vec x)|0_{\rm eff}(\eta) \rangle
\nonumber \\
&\!\!\!=\!\!\!&
- {ic_3^* m^3\over 48\pi^2} 
{{\cal G}_{\mu\nu,}^{\ \ ij}(\eta,\eta_0)\delta_{ij}\over a(\eta_0)} 
\biggl[ {1\over\epsilon} - \gamma + \ln {4\pi\mu^2\over m^2} \biggr] 
\nonumber \\
&&
+ {1\over a(\eta_0)} 
{\cal G}_{\mu\nu,}^{\ \ \lambda\rho}(\eta,\eta_0) 
\delta T_{\lambda\rho}(\eta_0)
+ \cdots . 
\label{eg3andct}
\end{eqnarray}
Thus, the structure needed to cancel the divergence is $\delta T_{00} = \delta T_{0i} = 0$ along with  
\begin{equation}
\delta T_{ij} = \delta_{ij} {ic_3^* m^3\over 48\pi^2} 
\biggl[ {1\over\epsilon} - \gamma + \ln {4\pi\mu^2\over m^2} \biggr] . 
\label{ct3eval}
\end{equation}

To understand the geometric origin of this counterterm, consider the action for surface tension, 
\begin{equation}
S_\sigma = \int d^3\vec x\, \sqrt{\hat g}\, \sigma . 
\label{surftension}
\end{equation}
The only appearance of the metric occurs in the determinant of the induced metric; to linear order in the fluctuation we have that 
\begin{equation}
\sqrt{\hat g} = a^3 \bigl( 1 + {\textstyle{1\over 2}} h^{ij} \delta_{ij} 
+ \cdots \bigr) 
\label{linearhatg}
\end{equation}
so that the leading contribution from this boundary action to the interaction Hamiltonian occurs through 
\begin{eqnarray}
S_\sigma &\!\!\!=\!\!\!& 
\int d^3\vec x\, a^3 
h^{ij} \bigl\{ {\textstyle{1\over 2}} \sigma \delta_{ij} \bigr\} 
+ \cdots 
\nonumber \\ 
&\!\!\!=\!\!\!& 
\int_{\eta_0}^0 d\eta \int d^3\vec x\, a^3 
h^{ij} \bigl\{ {\textstyle{1\over 2}} \sigma \delta_{ij} \bigr\} \, \delta(\eta-\eta_0)
+ \cdots \qquad 
\label{tensionh}
\end{eqnarray}
which matches with Eq.~(\ref{ct3eval}) when 
\begin{equation}
\sigma = {ic_3^* m^3\over 24\pi^2} 
\biggl[ {1\over\epsilon} - \gamma + \ln {4\pi\mu^2\over m^2} \biggr] . 
\label{sigmaval}
\end{equation}
Note that $c_3$ should be purely imaginary for the action to remain real; this behavior is exactly consistent with what was found when renormalizing an interacting scalar field in an effective initial state \cite{greens,dSgreens}.

\subsection{The extrinsic curvature}

The extrinsic curvature is the unique dimension one operator available on the boundary, so the divergences from the next moment in the initial state structure, 
\begin{equation}
f_k = c_2 {a^2(\eta_0) m^2\over\omega_k^2(\eta_0)} , 
\label{fk2eg}
\end{equation}
should be related---in part---to such a counterterm.  We now find that both the density and the pressure are divergent, 
\begin{eqnarray}
\rho_{\rm surf}(\eta') &\!\!\!=\!\!\!& 
- {c_2^* m^2\over 4} a^2(\eta_0) {a'(\eta')\over a^5(\eta')} 
K^{(2)\prime}(\eta')
+ \cdots 
\nonumber \\
p_{\rm surf}(\eta') &\!\!\!=\!\!\!& 
{c_2^* m^2\over 12} {a^2(\eta_0) \over a^4(\eta')} K^{(2)\prime\prime}(\eta') 
\nonumber \\
&& 
- {c_2^* m^2\over 4} a^2(\eta_0) {a'(\eta')\over a^5(\eta')} K^{(2)\prime}(\eta')
+ \cdots \qquad 
\label{rhopfk2} 
\end{eqnarray}
where we have again written the divergences in terms of derivatives of the kernel function.  This time, we have not explicitly shown any of the terms with logarithmic divergences since these are integrable.  Including these expressions in the graviton tadpole and integrating by part until we have isolated the kernel terms to the initial boundary, the divergent contributions from Eq.~(\ref{rhopfk2}) to the tadpole in Eq.~(\ref{surftadpole}) are 
\begin{eqnarray}
&&\!\!\!\!\!\!\!
\langle 0_{\rm eff}(\eta) | h_{\mu\nu}^+(\eta,\vec x)|0_{\rm eff}(\eta) \rangle
\nonumber \\
&\!\!\!=\!\!\!&
- {c_2^* m^2\over 8} {a'(\eta_0)\over a^3(\eta_0)} 
{\cal G}_{\mu\nu,}^{\ \ 00}(\eta,\eta_0)\, 
K^{(2)}(\eta_0)
\nonumber \\
&&
- {c_2^* m^2\over 8} {a'(\eta_0)\over a^3(\eta_0)} 
{\cal G}_{\mu\nu,}^{\ \ ij}(\eta,\eta')\delta_{ij}\, 
K^{(2)}(\eta_0)
\nonumber \\
&&
- {c_2^* m^2\over 24} a^2(\eta_0) {\partial\over\partial\eta'} 
\biggl[ {1\over a^4(\eta')} 
{\cal G}_{\mu\nu,}^{\ \ ij}(\eta,\eta')\delta_{ij} \biggr]_{\eta_0} 
K^{(2)}(\eta_0) 
\nonumber \\
&&
+ {c_2^* m^2\over 24} {1\over a^2(\eta_0)} 
{\cal G}_{\mu\nu,}^{\ \ ij}(\eta,\eta_0)\delta_{ij}\, K^{(2)\prime}(\eta_0) 
+ \cdots . \qquad
\label{surftadpolefk2}
\end{eqnarray}
Upon dimensionally regularizing the spatial momentum integrals, we extract the poles,  
\begin{eqnarray}
&&\!\!\!\!\!\!\!
\langle 0_{\rm eff}(\eta) | h_{\mu\nu}^+(\eta,\vec x)|0_{\rm eff}(\eta) \rangle
\nonumber \\
&\!\!\!=\!\!\!&
- {c_2^* m^2\over 96\pi^2} \biggl[ {1\over\epsilon} - \gamma + \ln 4\pi + \ln {\mu^2\over m^2} \biggr] 
\nonumber \\
&&\times 
\biggl[ 
3 {a'(\eta_0)\over a^3(\eta_0)} \bigl( 
{\cal G}_{\mu\nu,}^{\ \ 00}(\eta,\eta_0) 
+ {\cal G}_{\mu\nu,}^{\ \ ij}(\eta,\eta_0)\delta_{ij} \bigr) 
\nonumber \\
&&\quad
+ a^2(\eta_0) {\partial\over\partial\eta'} 
\biggl[ {1\over a^4(\eta')} 
{\cal G}_{\mu\nu,}^{\ \ ij}(\eta,\eta')\delta_{ij} \biggr]_{\eta_0} 
\biggr]
\nonumber \\
&&
+ \cdots . 
\label{DRtadpolefk2}
\end{eqnarray}
These divergent terms can be cancelled by adding the following surface counterterms, 
\begin{eqnarray}
H_I^{\rm c.t.} &\!\!\!=\!\!\!& 
\kappa \int d^3\vec x\, a^3(\eta) 
h^{\mu\nu} \biggl\{
{1\over a}\, \delta^i_\mu \delta^j_\nu \delta_{ij}\, \delta'(\eta-\eta_0) 
\nonumber \\
&&\qquad 
- {a'\over a^2}\, 
\bigl[ 3\delta^0_\mu \delta^0_\nu + \delta^i_\mu \delta^j_\nu \delta_{ij} \bigr]\, \delta(\eta-\eta_0) 
\biggr\} . \qquad  
\label{Hintfk2}
\end{eqnarray}
Notice that in addition to the $\delta$-function term that multiplies the dimension one term $a'/a^2$, we also have a $\delta$-function derivative.  This example illustrates the more general structure of the surface counterterms, which will consist most generally of products of derivatives of the scale factor multiplying derivatives of the $\delta$-function.  In this case, the counterterm contribution to the graviton tadpole is 
\begin{eqnarray} 
&&\kappa \biggl[ 3 {a'(\eta_0)\over a^3(\eta_0)}
\bigl[ 
{\cal G}_{\mu\nu,}^{\ \ 00}(\eta,\eta_0) 
+ {\cal G}_{\mu\nu,}^{\ \ ij}(\eta,\eta_0) \delta_{ij} \bigr] 
\nonumber \\
&&\quad 
+ a^2(\eta_0) {\partial\over\partial\eta'} \biggl[ {1\over a^4(\eta')} 
{\cal G}_{\mu\nu,}^{\ \ ij}(\eta,\eta') \delta_{ij} \biggr]_{\eta_0} \biggr] . 
\qquad
\label{tadctfk2}
\end{eqnarray}
Comparing Eqns.~(\ref{DRtadpolefk2}) and (\ref{tadctfk2}), we see that can cancel the divergence by choosing 
\begin{equation}
\kappa = {c_2^* m^2\over 96\pi^2} \biggl[ {1\over\epsilon} - \gamma + \ln 4\pi + \ln {\mu^2\over m^2} \biggr] . 
\label{choosekap}
\end{equation}

Since the initial state necessarily breaks some of the space-time symmetries, the strict relation between the standard geometric invariants on the surface and the form of the actual counterterms becomes more complicated beyond the simple case of a surface tension.  For example, if we choose the $\delta$-function derivative to match that needed, the contribution from the linear part of the extrinsic curvature is 
\begin{eqnarray}
&&\!\!\!\!\!\!\!\!\!\!\!\!\!\!\!\!\!\!\!\!\!\!
- 2\kappa \int d^3\vec x\, \sqrt{\hat g}\, K 
\nonumber \\
&\!\!\!=\!\!\!& 
- \int_{\eta_0}^0 d\eta\, 
\kappa \int d^3\vec x\, a^3 h^{\mu\nu} 
\biggl\{ 
{1\over a} \delta_\mu^i \delta_\nu^j \delta_{ij}\, \delta'(\eta-\eta_0) 
\nonumber \\
&&\quad
- {a'\over a^2} \bigl[ 
6 \delta_\mu^0 \delta_\nu^0 - 5 \delta_\mu^i \delta_\nu^j \delta_{ij} 
\bigr]\, \delta(\eta-\eta_0)
+ \cdots \biggr\} ;
\label{SKtoh}
\end{eqnarray}
so the extrinsic counterterm accounts for a part of the boundary action needed to cancel the $c_2$ moment of the initial state, 
\begin{eqnarray}
S_I^{\rm c.t.} &\!\!\!\equiv\!\!\!& 
- \int_{\eta_0}^0 d\eta\, H_I^{\rm c.t.}(\eta) 
\nonumber \\
&\!\!\!=\!\!\!&
- 2\kappa \int d^3\vec x\, \sqrt{\hat g}\, K 
+ \kappa \int d^3\vec x\, \sqrt{\hat g_{\rm cl}}\, K_{\rm cl}
\bigl[ h^{00} - 2 h \bigr] 
\nonumber \\
&&
+ \cdots . 
\label{SKtohwHI}
\end{eqnarray}
Here $\hat g_{\rm cl}$ and $K_{\rm cl}$ are evaluated for the purely classical limit ($h_{\mu\nu}\to 0$).

\subsection{A dimension two surface counterterm}

As a final example, let us consider an initial state whose structure is described by 
\begin{equation}
f_k = c_1 {a(\eta_0) m\over\omega_k(\eta_0)} . 
\label{fk1eg}
\end{equation}
As we add each new factor of the momentum to the short-distance structure, we obtain an additional derivative of the scale factor or the kernel in the terms that diverge at the boundary faster than $\ln(\eta-\eta_0)$, 
\begin{eqnarray}
\rho_{\rm surf}(\eta') &\!\!\!=\!\!\!& 
{ic_1^*m\over 8} a(\eta_0) \biggl\{ 
- {a'(\eta')\over a^5(\eta')} K^{(1)\prime\prime}(\eta') 
\nonumber \\
&& 
+ \biggl[ 
{a^{\prime\prime}(\eta')\over a^5(\eta')} 
+ {a^{\prime\, 2}(\eta')\over a^6(\eta')} \biggr] K^{(1)\prime}(\eta')
+ \cdots \biggr\} 
\qquad
\label{rhofk1}
\end{eqnarray}
and 
\begin{eqnarray}
p_{\rm surf}(\eta') &\!\!\!=\!\!\!& 
{ic_1^*m\over 24} a(\eta_0) \biggl\{ 
{1\over a^4(\eta')} K^{(1)\prime\prime\prime}(\eta') 
\nonumber \\
&&
- 3 {a'(\eta')\over a^5(\eta')} K^{(1)\prime\prime}(\eta') 
+ 2 {a^{\prime\prime}(\eta_0)\over a(\eta_0)} {1\over a^4(\eta')} K^{(3)\prime\prime}(\eta') 
\nonumber \\
&& 
- \biggl[ 
2 {m^2\over a^2(\eta')} 
+ {a^{\prime\prime}(\eta')\over a^5(\eta')} 
- 3 {a^{\prime\, 2}(\eta')\over a^6(\eta')} \biggr] K^{(1)\prime}(\eta')
\nonumber \\
&& 
+ \cdots \biggr\}
\label{pfk1}
\end{eqnarray}
If we insert these contributions of the energy-momentum in the tadpole condition of Eq.~(\ref{surftadpole}), integrate by parts a sufficient number of times to isolate the divergences on the initial boundary and then dimensionally regularize the remaining spatial momentum integrals, we obtain the following poles, 
\begin{eqnarray}
\hbox{Eq.~(\ref{surftadpole})} &\!\!\!=\!\!\!& 
{ic_1^*m\over 96\pi^2} \biggl[ 
{1\over\epsilon} - \gamma + \ln 4\pi + \ln {\mu^2\over m^2} \biggr] 
\nonumber \\
&& 
\times\biggl\{ 
{1\over a^3(\eta_0)} {\partial^2\over\partial\eta^{\prime 2}} 
{\cal G}_{\mu\nu,}^{\ \ ij}(\eta,\eta')\bigr|_{\eta'=\eta_0} \delta_{ij}
\nonumber \\
&& 
+ {a'(\eta_0)\over a^4(\eta_0)} {\partial\over\partial\eta'} 
\bigl[ 
{\cal G}_{\mu\nu,}^{\ \ 00}(\eta,\eta') 
- 7 {\cal G}_{\mu\nu,}^{\ \ ij}(\eta,\eta') \delta_{ij}
\bigr]_{\eta_0} 
\nonumber \\
&& 
+ \biggl[ 4 {a^{\prime\prime}(\eta_0)\over a^4(\eta_0)} 
- 2 {a^{\prime 2}(\eta_0)\over a^5(\eta_0)} \biggr]
{\cal G}_{\mu\nu,}^{\ \ 00}(\eta,\eta_0) 
\nonumber \\
&& 
- \biggl[ 4 {a^{\prime\prime}(\eta_0)\over a^4(\eta_0)} 
- 19 {a^{\prime 2}(\eta_0)\over a^5(\eta_0)} \biggr]
{\cal G}_{\mu\nu,}^{\ \ ij}(\eta,\eta_0) \delta_{ij}
\biggr\}
\nonumber \\
&& 
+ {\rm finite} . 
\label{polesfk1}
\end{eqnarray}
These divergences are cancelled by adding the following two-derivative term to the boundary
\begin{eqnarray}
S_\lambda^{\rm c.t.} &\!\!\!=\!\!\!& 
\lambda \int d^3\vec x\, 
\biggl\{ 
a \partial_0^2 h^{00} - a \partial_0^2 h 
+ 2 a' \partial_0 h^{00} - a' \partial_0 h 
\nonumber \\
&&\qquad\qquad
+ 4 a^{\prime\prime} h^{00} 
+ 5 {a^{\prime 2}\over a} h^{00} 
- 3 {a^{\prime 2}\over a} h 
\biggr\} , 
\label{fk1action}
\end{eqnarray}
where 
\begin{equation}
\lambda = - {ic_1^*m\over 96\pi^2} \biggl[ 
{1\over\epsilon} - \gamma + \ln 4\pi + \ln {\mu^2\over m^2} \biggr] . 
\label{lambdaval}
\end{equation}
In terms of the interaction Hamiltonian, this counterterm action yields 
\begin{eqnarray}
H_I^{\rm c.t.} &\!\!\!=\!\!\!& 
- \lambda \int d^3\vec x\, a^3 h^{\mu\nu} 
\biggl\{ {1\over a^2} \delta_\mu^i \delta_\nu^j \delta_{ij} \delta^{\prime\prime}(\eta-\eta_0) 
\nonumber \\
&&\qquad\qquad\qquad
- {a'\over a^3} \bigl[ \delta_\mu^0 \delta_\nu^0 - \delta_\mu^i \delta_\nu^j \delta_{ij} \bigr] \delta'(\eta-\eta_0) 
\nonumber \\
&&\qquad\qquad\qquad 
+ \biggl[ 3 {a^{\prime\prime}\over a^3} + 2 {a^{\prime 2}\over a^4} \biggr] \delta_\mu^0 \delta_\nu^0 \delta(\eta-\eta_0) 
\nonumber \\
&&\qquad\qquad\qquad 
+ 3 {a^{\prime 2}\over a^4} \delta_\mu^i \delta_\nu^j \delta_{ij}  \delta(\eta-\eta_0) \biggr\} . 
\label{fk1HI}
\end{eqnarray}

From the pattern shown in these examples, we can readily generalize to the case of the trans-Planckian terms---those scaling as some positive power of $\omega_k$ in Eq.~(\ref{fkdef}).  For the term 
\begin{equation}
d_n \biggl( {\omega_k(\eta_0)\over a(\eta_0) M} \biggr)^n 
\label{dneg}
\end{equation}
the counterterms, as part of the interaction Hamiltonian, will contain terms with $n+3$ derivatives which for $n>0$ represent irrelevant terms, according to the power counting applied for the three-dimensional boundary action.  The $n=0$ term is intermediate between the long and short-distance modifications of the initial state and accordingly requires a marginal counterterm.

\section{Conclusions}
\label{conclude}

The effective theory of an initial state provides a formalism for describing the signatures of new physics in a state that cannot be captured by an ordinary effective field theory.  When we apply this effective theory to inflation, this new physics can potentially refer to the dynamics at the Planck scale, where the correct theory of quantum gravity has yet to be established.  We have therefore allowed very general structures in the initial state, including those that do not obey the usual Hadamard condition at short distances.\footnote{For example, such non-Hadamard states can be useful for modeling the effects of a modification of the usual uncertainty relation at short-distances \cite{gary} or the appearance of some noncommutativity of space-time \cite{smolin}.}  So before we can apply this formalism, we should first show that it is perturbatively stable in both its field theoretical and gravitational components---that is, that the new short distance structures we have allowed in the state do not lead to any divergences that we cannot cancel though a suitable renormalization prescription.  Moreover, even after renormalizing the divergent parts of energy-momentum tensor, it is important to show that the remaining trans-Planckian contribution to the gravitational equations of motion, called the back-reaction, should be sufficiently small that it does not overwhelm the original vacuum energy that sustained the inflationary expansion \cite{schalm2,backreact,simeone}.  

In two earlier articles \cite{greens,dSgreens} we addressed the field theory side of the boundary renormalization.  In an interacting scalar theory, the loop corrections sum over the short-distance structures of the state; the new divergences associated with these structures are confined to the initial-time boundary and are removed by adding counterterms at that surface.  Similarly, the energy-momentum in a free scalar field is also sensitive to this initial state information and it also produces divergences at the initial-time boundary that are proportional to powers of the scale factor and its derivatives.  Here we have shown how these divergences are cancelled by adding gravitational counterterms at the boundary.  An interesting property of these counterterms is that they are relevant or irrelevant according to whether the corresponding moment of the initial state affects the state at long or at short distances.  The size of the back-reaction from the trans-Planckian part of the state was derived in \cite{emtensor} and was shown to be never larger than $(H^2/M^2_{\rm pl})(H/M)$, viewed as a fraction of the inflationary vacuum energy.

Once we have a reliable framework, we can begin to examine how the small corrections from phenomena above the Hubble scale can influence the primordial power spectrum predicted by inflation.  It is important to have such a formalism since the power spectrum will continue to be measured with an increasing accuracy throughout the next decade or so.  Current \cite{wmap} and future \cite{planck} experiments that measure the fluctuations in the cosmic microwave background should eventually achieve an accuracy at the level of one part in $10^3$.  Future galaxy surveys \cite{ska}, by following the distribution of structures in the universe before the nonlinearities of their gravitational collapse have begun, should be able to measure the power spectrum to about one part in $10^5$, approaching the theoretical limit of one part in $10^6$ \cite{spergel}.  To provide some perspective, it is useful to recall that the acoustic oscillations seen by WMAP themselves represent roughly a one part in $10^5$ perturbation to an otherwise perfect blackbody spectrum.

The next stage of this work \cite{next} will be to determine the leading trans-Planckian signal in the microwave background radiation.  By adding two extra parameters\footnote{In the power series expansion given in Eq.~(\ref{fkdef}) these parameters would be $d_1$ and a phase, which is related to $\eta_0$.} to the minimal set used to fit the observed acoustic oscillations, we can constrain the possible trans-Planckian component of the initial state.  From previous studies \cite{ekp}, the trans-Planckian signature tends to be rather distinctive, with a correlated amplitude and period in the correction to the primordial power spectrum.  This correlation does not naturally result, for example, from features in the inflaton potential, so it should be possible to distinguish the signatures for these two sources for small corrections to the power spectrum---the inflaton potential and the effective state.

The advantage of any effective theory is that it depends only very generally on the assumed symmetries of a system and is rather independent of the details of a particular model.  So once we have fixed the leading parameter of the effective state observationally, we can apply this constraint to any model by expanding its mode functions in a power series resembling that in Eq.~(\ref{fkdef}) and extracting the coefficient of the leading trans-Planckian term.  In fact, many of the past models were originally defined in terms of the behavior of their mode functions, so translating between the effective theory and specific models for the trans-Planckian structure should be fairly straightforward.  In future work we shall examine this connection between the effective theory and particular models, using a composite inflaton field as an detailed test case.

\begin{acknowledgments}

\noindent
This work was supported in part by DOE grant No.~DE-FG03-91-ER40682 and the National Science Foundation grant No.~PHY02-44801.  

\end{acknowledgments}


\appendix 

\section{Kernels}
\label{kernels}

Because the divergences associated with the effective initial state both occur specifically at the boundary and moreover depend on the specific details of the state, it is useful to introduce a general kernel function which is applicable to any boundary divergence.  In Sec.~\ref{graviton} we found that a logarithmically divergent integrand can be integrated to yield a finite contribution to the graviton tadpole.  Therefore we define the kernel function for this threshold case, 
\begin{equation}
K^{(p)}(\eta) = \int {d^3\vec k\over (2\pi)^3}\, 
{e^{-2i\int_{\eta_0}^\eta d\eta'\, \Omega_k(\eta')}\over \Omega_k^{3-p}(\eta) \Omega_k^p(\eta_0) } , 
\label{pkernel}
\end{equation}
so that $K^{(p)}(\eta_0)$ has a logarithmic divergence from the large $k$ region of the integral.  We can easily generate an arbitrary divergent momentum integral by differentiating $K^{(p)}(\eta)$ a sufficient number of times; among the examples we have presented, we encountered linearly and quadratically divergent contributions to the tadpole integrand which are produced by 
\begin{eqnarray}
K^{(p)\prime}(\eta) &\!\!\!=\!\!\!& 
- 2i \int {d^3\vec k\over (2\pi)^3}\, 
{e^{-2i\int_{\eta_0}^\eta d\eta'\, \Omega_k(\eta')}\over \Omega_k^{2-p}(\eta) \Omega_k^p(\eta_0) } 
\nonumber \\
&&\qquad\qquad\times 
\biggl\{ 
1 - {i\over 2} (3-p) {\Omega'_k(\eta)\over \Omega_k^2(\eta) }
\biggr\}
\label{Dpkernel}
\end{eqnarray}
and 
\begin{eqnarray}
K^{(p)\prime\prime}(\eta) &\!\!\!=\!\!\!& 
- 4 \int {d^3\vec k\over (2\pi)^3}\, 
{e^{-2i\int_{\eta_0}^\eta d\eta'\, \Omega_k(\eta')}\over\Omega_k^{1-p}(\eta) \Omega_k^p(\eta_0) } 
\nonumber \\
&&\quad\times 
\biggl\{ 
1 - {i\over 2} (5-2p) {\Omega'_k(\eta)\over\Omega_k^2(\eta) }
+ {1\over 4} (3-p) {\Omega^{\prime\prime}_k(\eta)\over\Omega_k^3(\eta) }
\nonumber \\
&&\quad\quad
- {1\over 4} (3-p)(4-p) {\bigl( \Omega'_k(\eta) \bigr)^2\over\Omega_k^4(\eta)}
\biggr\} . 
\label{DDpkernel}
\end{eqnarray}
Since we are not concerned with the finite parts of the integrand of the tadpole---these just modify slightly the standard ``bulk'' equations of motion---we expand these general kernels in terms of the frequency $\omega_k$ we introduced for our adiabatic approximation.  Expanding each of the three cases above and only retaining the terms that diverge at $\eta=\eta_0$, we find 
\begin{equation}
K^{(p)}(\eta) = \int {d^3\vec k\over (2\pi)^3}\, 
{e^{-2i\int_{\eta_0}^\eta d\eta'\, \Omega_k(\eta')}\over \omega_k^{3-p}(\eta) \omega_k^p(\eta_0) } 
\biggl\{ 1 + {\cal O} \biggl( {1\over\omega_k^2} \biggr) \biggr\}
\label{pkernelo}
\end{equation}
and 
\begin{equation}
K^{(p)\prime}(\eta) = 
- 2i \int {d^3\vec k\over (2\pi)^3}\, 
{e^{-2i\int_{\eta_0}^\eta d\eta'\, \Omega_k(\eta')}\over \omega_k^{2-p}(\eta) \omega_k^p(\eta_0) } 
\biggl\{ 1 + {\cal O} \biggl( {1\over\omega_k^2} \biggr) \biggr\}
\label{Dpkernelo}
\end{equation}
and 
\begin{eqnarray}
K^{(p)\prime\prime}(\eta) &\!\!\!=\!\!\!& 
- 4 \int {d^3\vec k\over (2\pi)^3}\, 
{e^{-2i\int_{\eta_0}^\eta d\eta'\, \Omega_k(\eta')}\over\omega_k^{1-p}(\eta) \omega_k^p(\eta_0) } 
\nonumber \\
&&\quad\times 
\biggl\{ 
1 + {1\over 2} {1-p\over\omega_k^2(\eta)} 
{a^{\prime\prime}(\eta)\over a(\eta)}
+ {1\over 2} {p\over\omega_k^2(\eta_0)} 
{a^{\prime\prime}(\eta_0)\over a(\eta_0)}
\nonumber \\
&&\qquad
+ {\cal O} \biggl( {1\over\omega_k^4} \biggr) 
\biggr\} . 
\label{DDpkernelo}
\end{eqnarray}

\end{document}